\documentclass{aa}  

\usepackage{graphicx}
\usepackage{txfonts}
\usepackage{subcaption}     
\usepackage{natbib}    
\usepackage{xcolor}
\usepackage{tikz}
\usepackage{todonotes}
\usepackage{amsmath}
\usepackage{placeins}

\providecommand{\comment}[1]{}
\renewcommand{\comment}[1]{\textcolor{red}{#1}}
\newcommand{\cob}{CO$^{5}$BOLD}

\newcommand{\Rsol}{\ensuremath{R_\odot}}
\newcommand{\mjup}{\ensuremath{M_{\mathrm{Jup}}}}
\newcommand{\msol}{\ensuremath{M_{\odot}}}
\newcommand{\lsol}{\ensuremath{L_{\odot}}}

\usepackage{hyperref}
\hypersetup{colorlinks=true,allcolors=[rgb]{0,0,1}}

\begin{document}

   \title{AGB Atmospheres in Binary Environments}

   \subtitle{A First Step with \cob}

%

   \author{G. Tomassini \inst{1}
           \and A. Chiavassa \inst{1,2}
           \and B. Freytag \inst{3}
           \and E. Lagadec \inst{1}
          }

   \institute{Université Côte d'Azur, Observatoire de la Côte d'Azur, CNRS, Laboratoire Lagrange, France
   \and
    Max Planck Institute for Astrophysics, Karl-Schwarzschild-Str. 1, 85748 Garching, Germany;
   \and
   Theoretical Astrophysics, Department of Physics and Astronomy, Uppsala University, Box 516, SE-751 20 Uppsala, Sweden
   }

   \date{}
 
  \abstract
   {Asymptotic Giant Branch (AGB) stars are characterized by complex atmospheric dynamics driven by convection and pulsations. In binary systems, the presence of a companion can further influence these dynamics, potentially affecting mass-loss processes and the morphology of the circumstellar environment, yet this influence on the atmosphere itself remains poorly quantified.
   }
   {This study aims to fill this gap by employing \cob\ simulations with a modified gravitational potential to explore how companion mass affects the atmospheric structure of AGB stars. We seek to elucidate the role of binarity in shaping AGB atmospheres and its implications for mass loss processes.}
    {We used radiative hydrodynamics (RHD) simulations with \cob\ to model the convective interior and atmosphere of AGB stars. We modified the gravitational potential to include a secondary mass and a centrifugal term, representing the companion as a point-mass potential on a steady, tidally locked circular orbit outside the computational domain, with the Coriolis force neglected. We performed a series of simulations with different companion masses to investigate their effects on the AGB atmosphere. Using singular value decomposition (SVD), we identified the spatial distribution of pulsation energy and quantified its properties. We also examined the density structure to identify anisotropies induced by the companion.}
   {This SVD decomposition isolates two dynamically distinct regimes, a pulsation-dominated layer and an outer, low-frequency outflow region, whose pulsation properties remain largely unaffected by the companion: their frequency, amplitude, and coherence stay consistent across all models. The density structure, by contrast, responds clearly to the companion, forming a pronounced anisotropy along the binary axis that strengthens with companion mass. This indicates that the intrinsic pulsational dynamics of the AGB star dominate over the tidal influence of the companion within the explored parameter space, while the density structure remains the most sensitive tracer of binarity.}
   {}


   \maketitle
\nolinenumbers
\section{Introduction}

Asymptotic Giant Branch (AGB) stars mark the final nuclear-burning stage in the evolution of low- and intermediate-mass stars ($\sim 0.8-8\,M_\odot$). Once core helium burning ceases, a degenerate carbon--oxygen core is left, surrounded by a helium-burning and a hydrogen-burning shell. The helium shell burns unstably in recurrent thermal pulses, and the star spends a long time on this thermally pulsing AGB \citep{Herwig2005}. Such a star has a deep, strongly convective envelope that pulsates with large amplitude, surmounted by an extended, cool atmosphere that is mostly convectively stable and follows the pulsations of the star below. These pulsations produce shock waves in the atmosphere that lift the gas to radii where it is cool enough for dust to condense. Radiation pressure on the dust then accelerates the gas outward, and this dust-driven wind progressively strips the envelope \citep{HofnerOlofsson2018}. Eventually the core is exposed as the star evolves into a post-AGB (pAGB) object and, ultimately, a planetary nebula. This late evolutionary sequence (AGB and pAGB) is governed by intense mass loss, with rates of order $10^{-7}-10^{-5}\,M_\odot\,\mathrm{yr}^{-1}$ \citep{Knapp1998, Ramstedt2009}. While the dust-driven wind scenario explains the bulk of the mass loss, the origin of the asymmetries observed in the circumstellar envelopes of AGB stars (spirals, arcs, discs, and bipolar outflows) remains debated, and the role of a binary companion in shaping the wind close to the star is still poorly constrained \citep{HofnerOlofsson2018,Decin2020}. Beyond the evolution of the star itself, a growing body of evidence indicates that binarity can substantially alter this picture at essentially every stage of the AGB and post-AGB evolution \citep{DeMarco2009}.

Indeed, observational evidence accumulated over the last decade shows that binarity plays a pivotal role in shaping late stellar evolution: an estimated $\sim 30\%$ of post-AGB stars are found in binaries with periods of 100-1500 days, and roughly one-fifth of AGB envelopes with high mass-loss rates already display morphological signatures attributable to a companion \citep{DeMarco2009}. Post-AGB systems frequently exhibit circumbinary disks, tidal structures, and asymmetric outflows, and spiral, arc, and equatorial density-enhancement patterns have been directly observed in the winds of several evolved stars and attributed to the gravitational influence of a (sub)stellar companion \citep{MauronHuggins2006,KimTaam2012,Decin2020, Badolo2026}. Hydrodynamic studies have further shown how companions shape this ejected material on large scales, well beyond the stellar surface \citep{Mastrodemos1999,Liu2017,Chen2018,ElMellah2020,Decin2020,aydi2022}. However, the impact of a companion on the AGB atmosphere itself -- the region where the wind is actually launched -- remains poorly understood, largely because it requires resolving the interplay between convection, pulsation, and the companion's gravitational field simultaneously, which large-scale wind simulations do not capture.

Even without a companion, this interplay is already a significant source of complexity. Global three-dimensional radiation-hydrodynamics simulations show that convection and stellar pulsations self-consistently generate shocks and large-scale asymmetries in the atmospheres of single AGB stars \citep{Freytag2023}, with direct consequences for the emerging stellar structure. Introducing a companion -- whether of planetary or stellar mass -- into this already dynamically active atmosphere could plausibly enhance or reorganize these asymmetries, but the extent of such an influence has yet to be quantified.

A key observational obstacle is that companions of AGB stars are typically undetectable directly: they are faint, and embedded within the dusty and dynamically active circumstellar environment of their host. The nearby AGB star L$_2$ Puppis (L$_2$ Pup, $d = 64$\,pc) is an exception to this rule, and offers a rare opportunity to calibrate this study against an object with an actually measured companion. High-contrast imaging has revealed an edge-on circumstellar dust disk together with bipolar ''hourglass'' cones and collimated streamers around the star \citep{Kervella2014,Kervella2015a}. This morphology is classically associated with the hydrodynamical imprint of a low-mass companion \citep{Nordhaus2006}. Numerical simulations of L$_2$ Pup as a binary system have been able to reproduce \citep{Chen2016} this morphology. Crucially, because the companion is embedded in a resolved molecular disk, the very high angular resolution of ALMA made it possible to go beyond morphology and directly measure its dynamical mass and orbit. \cite{Kervella2016} constrained L$_2$~Pup B to a mass of $m_B = 12 \pm 16\,\mjup$, orbiting at a true radius $R_B = 2.43 \pm 0.16$\,AU (period $P_{\mathrm{orb}} \approx 4.7$\,yr) around a primary of mass $M_A = 0.659 \pm 0.043\,\msol$. L$_2$ Pup therefore stands as direct observational evidence that planetary- to brown-dwarf-mass companions can orbit within a few stellar radii of an evolved giant, and it provides a physically motivated benchmark of companion mass and separation for this study\footnote{The evolutionary status of L$_2$ Pup A itself is debated: \cite{Uttenthaler2024} recently argued, from its Tc-poor composition and pulsation properties, that the star is statistically more likely to be on the RGB than genuinely on the AGB. This does not affect its relevance as a benchmark for the companion's mass and orbit, but calls for some caution when using L$_2$ Pup as an AGB prototype.}.

To model the atmospheric response to such a companion, we use \cob\ \citep{Freytag2012,Freytag2017a}, a three-dimensional ``star-in-a-box'' radiation-hydrodynamics code that self-consistently couples compressible hydrodynamics with non-local radiative energy transfer. This combination of hydrodynamics and radiative transfer, solved together rather than parametrised, makes \cob\ particularly well suited to modelling AGBs, where convection, pulsation, and the shocks they generate emerge directly from the simulated physics and jointly determine the emerging stellar structure. In its standard single-star configuration, \cob\ can be tuned to fundamental parameters comparable to those of the L$_2$ Pup system. Such single-star models use the stellar mass and luminosity as input parameters, whereas radius and effective temperature are affected by the atmospheric dynamics and can be determined only after a simulation has finished \citep{ Ahmad2023}. The convective surface patterns also displace the photocentre, which affects astrometric observables such as Gaia parallaxes \citep{Beguin2024}. This study extends that same modelling framework to include, for the first time, the dynamical effect of a low-mass companion, using the L$_2$ Pup system to motivate and anchor the explored range of companion masses and separations.

This study aims to determine what observable imprint a companion in the L$_2$ Pup mass and separation range leaves on the atmosphere of an AGB star, and whether such a signal can be disentangled from the intrinsic stochastic variability of convection. We employ modified \cob\ simulations, in which the stellar gravitational potential is extended to include a secondary mass and the associated centrifugal term (Sect.~\ref{Sec:Methods}), to explore how companions of varying mass and orbital separation affect the atmospheric structure of AGB stars. By systematically varying these parameters across a grid motivated by the parameters of the L$_2$ Pup system, we seek to elucidate the role of binarity in shaping AGB atmospheres and its implications for the initiation of mass loss.

\section{Methods}
\label{Sec:Methods}

\subsection{General setup}

We employ three-dimensional Radiative Hydrodynamics (RHD) simulations with the \cob\ code \citep{Freytag2012, Freytag2017a} to model the convective interior and atmosphere of AGB stars. \cob\ solves the coupled, non-linear equations of compressible hydrodynamics -- using an approximate Roe solver -- together with non-local radiative energy transfer, computed with a short-characteristics method, on a Cartesian grid of cubical cells. In this central volume, heat is added as a constant source term, corresponding to the stellar luminosity. A drag force is active in this core region only, to limit the amplitude of dipolar flows traversing the entire star. All outer boundaries are open for the flow of matter and for radiation (see \cite{Freytag2017a}, for some details about boundary conditions in CO5BOLD). This setup self-consistently captures the dynamics of the outer convective interior and the atmosphere, where shocks triggered by internal pulsations drive the stellar stratification \citep{Freytag2023}.

Grey Rosseland gas opacities are computed from COMA opacity tables \citep{Aringer2000, Aringer2016}, assuming solar abundances $(X, Y, Z) = (0.7088, 0.2750, 0.0162)$. Although \cob\ can include time-dependent dust formation, the models presented in this work are dust-free.

\subsection{Modified potential}

To simulate the binary system, we modify the standard time-independent, spherically symmetric gravitational potential of \cob\ to include a secondary mass and the associated centrifugal term. The primary star, located at position $\mathbf{r}_\star$, already uses a smoothed monopole potential of the form

\begin{equation}
   \Phi_{\star}(\mathbf{r}) = -GM_{\star} \left[ r_0^4 + \frac{|\mathbf{r}-\mathbf{r}_\star|^4}{\sqrt{1+\left(|\mathbf{r}-\mathbf{r}_\star|/r_1\right)^8}} \right]^{-1/4} \enspace,
\end{equation}

\noindent where $r_0$ is a softening length that regularises the potential at the centre of the computational domain, and $r_1$ is an outer smoothing radius: for $r_0 \ll |\mathbf{r}-\mathbf{r}_\star| \ll r_1$, $\Phi_\star$ follows the standard point-mass potential $-GM_\star/|\mathbf{r}-\mathbf{r}_\star|$, while beyond $r_1$ it smoothly levels off, $\Phi_\star \xrightarrow[|\mathbf{r}-\mathbf{r}_\star| \to \infty]{} -GM_\star/r_1$. This artificially reduces gravity in the outermost layers (the corners of the box), increasing the pressure and density scale heights there.

We introduce a companion of mass $M_{\mathrm{c}}$ at position $\mathbf{r}_{\mathrm{c}}$, at a separation $d = |\mathbf{r}_{\mathrm{c}} - \mathbf{r}_\star|$ from the primary, by adding an analogous term to the potential:

\begin{equation}
   \Phi_{\mathrm{c}}(\mathbf{r}) = -GM_{\mathrm{c}} \left[ r_0^4 + \frac{|\mathbf{r}-\mathbf{r}_{\mathrm{c}}|^4}{\sqrt{1+\left(|\mathbf{r}-\mathbf{r}_{\mathrm{c}}|/r_1\right)^8}} \right]^{-1/4} \enspace.
\end{equation}

In addition, we include a centrifugal term to account for the orbital motion of the binary system,

\begin{equation}
   \Phi_{\mathrm{cent}}(\mathbf{r}) = -\frac{1}{2} \left| \mathbf{\Omega} \times \mathbf{r} \right|^2\enspace \enspace,
\end{equation}

\noindent where $\mathbf{\Omega} = \sqrt{G(M_{\star} + M_{\mathrm{c}})/d^3}\, \mathbf{\hat{z}}$ is the angular velocity for a separation $d$, directed along the rotation axis $\mathbf{\hat{z}}$ of the orbit. Therefore, the total effective potential in the corotating frame of the binary system is

\begin{equation}
   \Phi_{\mathrm{eff}}(\mathbf{r}) = \Phi_{\star}(\mathbf{r}) + \Phi_{\mathrm{c}}(\mathbf{r}) + \Phi_{\mathrm{cent}}(\mathbf{r}) \enspace,
\end{equation}

\begin{figure}[h]
   \centering
   \subfloat[Schematic representation of the binary system]{
   \includegraphics[width=0.4\textwidth]{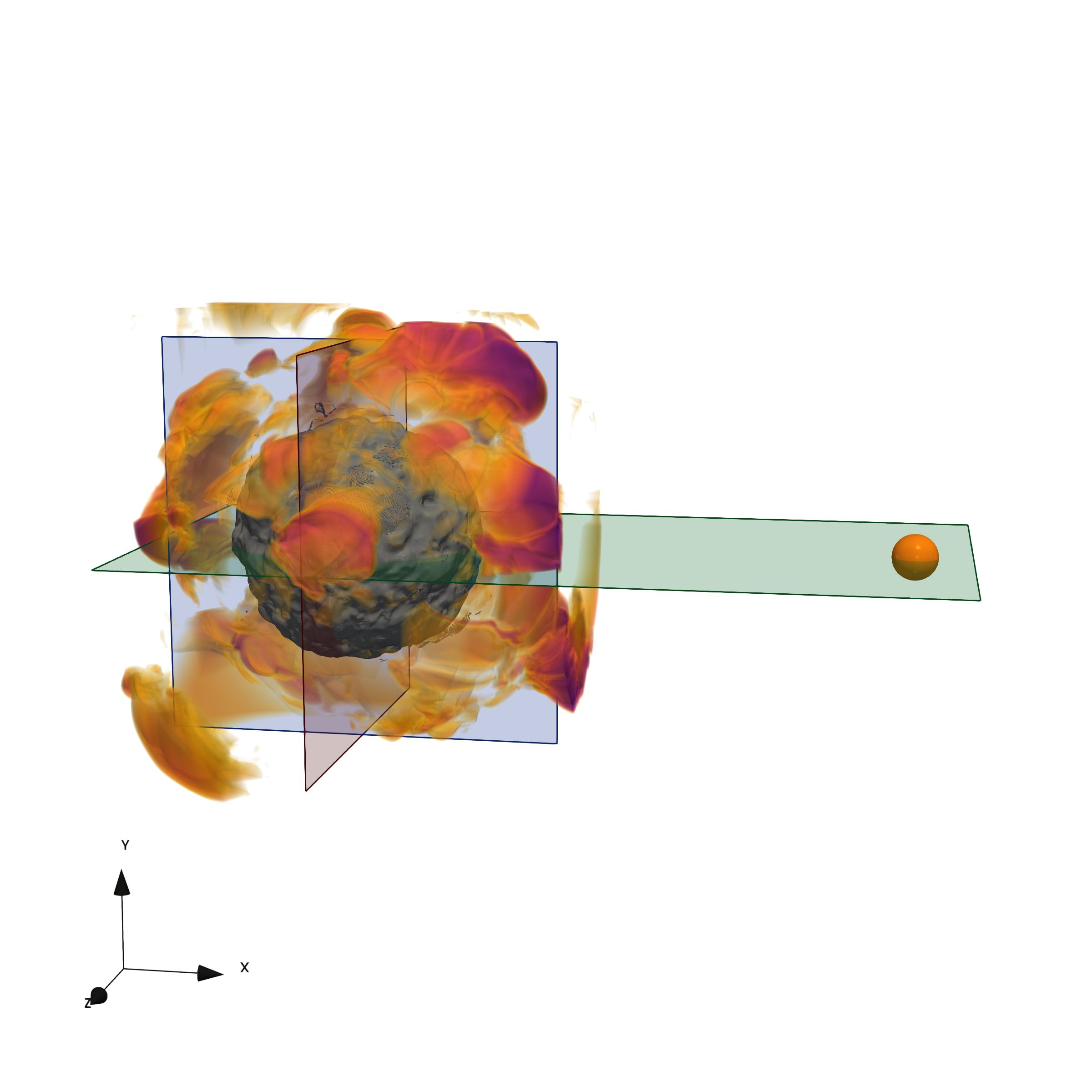}
   \label{fig:potential_binary_scheme}
   }\\
   \subfloat[Contour plot of the effective potential]{
   \includegraphics[width=0.5\textwidth]{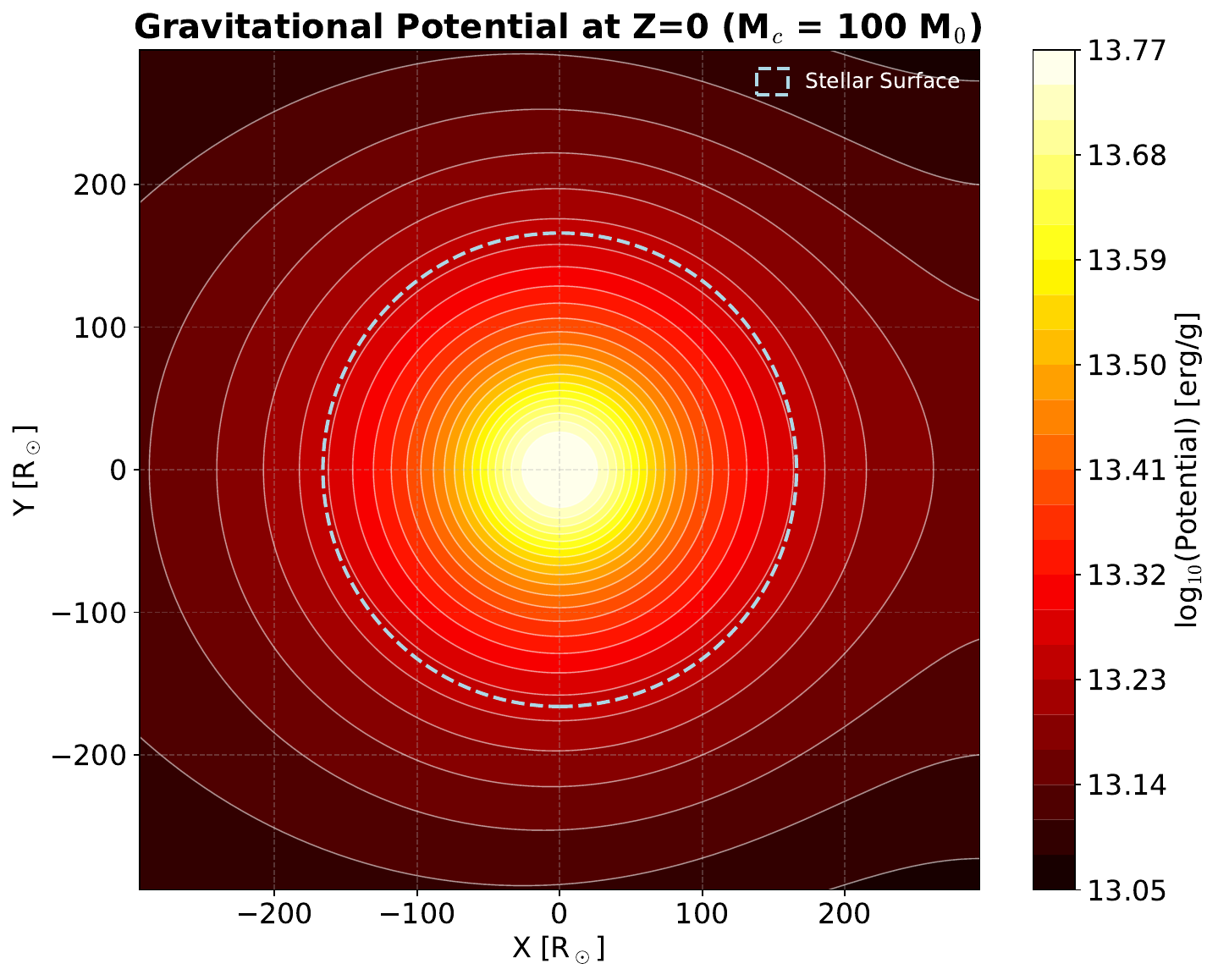}
   \label{fig:potential_binary_contour}
   }
   \caption{(\textbf{top}) Schematic representation of the modified gravitational potential in a binary system. The primary star (AGB) is located at the origin, while the companion is positioned at a distance $d$ along the x-axis. The grey "sphere" represents an isothermal surface at $T = 4000$\,K, and the colormap is the Mach number computed from the radial velocity (only Mach > 0 are shown). (\textbf{bottom}) Contour plot of the effective potential. The effective potential $\Phi(\mathbf{r})$ combines the gravitational influences of both masses and the centrifugal term due to orbital motion.}
   \label{fig:potential_binary}
\end{figure}

\noindent This setup assumes a steady circular orbit and tidal locking, neglecting orbital evolution. A schematic representation of the setup and a 2D contour plot of the modified potential are shown in Figure~\ref{fig:potential_binary}.

\subsection{Simulation grid}
The primary star is an AGB model with $M_{\star} = 1.0 M_{\odot}$ and $T_{\mathrm{eff}} = 3160$\,K ($R_{\star} \approx 166\,R_{\odot}$), within a box of $\sim (\,3.5 \,R_{\star})^3$. 

Our binary grid is anchored on the parameters of L$_2$~Pup B \citep{Kervella2016}. To maintain a consistent tidal impact while ensuring that the companion remains outside the computational domain, we set the separation to $d = 3.34$\,AU ($\sim 4.3\,R_{\star}$). For simplicity, the companion is modelled as a point mass on the x-axis, so $\mathbf{r_c} = (d, 0, 0)$. We parametrise the companion mass $M_c$ through the tidal scale height $h$ (see Appendix~\ref{Sec:tide}):

\begin{equation}
    \label{eq:tide}
   h = \frac{3 M_c R_\star^4}{2 M_\star d^3}
   \enspace ,
\end{equation}

\noindent with $M_\star = 1\,M_\odot$. Since the tidal effect depends on the height relative to the stellar radius, we scale our lowest-mass model on $h/R_\star$: for L$_2$~Pup~B, $h \simeq 0.04\,R_\odot$ corresponds to $h/R_{\rm L_2\,Pup} = 0.034\,\%$, and we adopt $h_0 = 0.05\,R_\odot$ for our simulation, i.e. $h_0/R_\star = 0.032\,\%$, within the uncertainty on the mass of L$_2$~Pup~B. Inverting the equation above gives the corresponding companion mass $M_0 \simeq 18\,\mjup$. The other models of the grid use masses that are multiples of $M_0$, $M_c = n\,M_0$, from $n=1$ ($18\,\mjup$) to $n=100$ ($1820\,\mjup$). We do not go below $M_0$, because the tidal effect is already very small there ($h_0/R_\star \approx 3\times10^{-4}$)..

The full grid is summarized in Table~\ref{tab:models}. A time series of 3D snapshots of the Mach number for each model can be found in Figure~\ref{fig:snapshots}. The color scale represents the Mach number, i.e. the ratio between the local velocity and the sound speed, with warmer colors indicating higher speeds (\cite{Chiavassa2024}, Figure 6). The grey sphere represents the isothermal surface at $T = 4000$\,K, which corresponds to the temperature where $\tau = 33$. We chose this temperature to probe deeper layer than $\tau =1$ to avoid having extended convective cells in the snapshots.

\begin{table*}[t]
   \centering
   \caption{Model parameters for the binary system simulations.}
   \footnotesize
   \setlength{\tabcolsep}{5pt}
   \begin{tabular}{lrrrrrrrrr}
      \hline\hline
   Model & $M_1$ [\msol] & $M_2$ [\mjup] & $q$ & $h/R_\star$ [\%] & $L$ [\lsol] & $T_{\mathrm{eff}}$ [K] & $\log g$ & $N_t$ & $dx$ [\Rsol] \\
   \hline
   L$_2$ Pup & $0.659\pm0.011$ & $12\pm16$ & 0.017 & 0.034 & $2000\pm700$ & $3500\pm250$ & $0.078\pm0.027$ & -- & -- \\
   \hline
   ref       & 1 & --   & --    & --     & $2476\pm59$ & $3164\pm19$ & $-0.004\pm0.004$ & 6525 & 1.75 \\
   m1        & 1 & 18   & 0.017 & 0.032 & $2496\pm76$ & $3165\pm25$ & $-0.006\pm0.005$ & 6525 & 1.75 \\
   m2        & 1 & 36   & 0.035 & 0.063 & $2484\pm67$ & $3155\pm25$ & $-0.010\pm0.008$ & 9585 & 1.75 \\
   m5        & 1 & 91   & 0.087 & 0.159 & $2486\pm72$ & $3155\pm26$ & $-0.010\pm0.008$ & 9345 & 1.75 \\
   m7\_5     & 1 & 136  & 0.130 & 0.238 & $2500\pm73$ & $3165\pm24$ & $-0.007\pm0.006$ & 7890 & 1.75 \\
   m10       & 1 & 182  & 0.173 & 0.317 & $2492\pm67$ & $3161\pm23$ & $-0.008\pm0.006$ & 7500 & 1.75 \\
   m15       & 1 & 272  & 0.255 & 0.476 & $2495\pm72$ & $3163\pm25$ & $-0.007\pm0.006$ & 9015 & 1.75 \\
   m50       & 1 & 910  & 0.868 & 1.587 & $2493\pm65$ & $3156\pm23$ & $-0.011\pm0.006$ & 6015 & 1.75 \\
   m100      & 1 & 1820 & 1.735 & 3.175 & $2497\pm60$ & $3152\pm22$ & $-0.014\pm0.006$ & 6015 & 1.75 \\
   \hline
   \end{tabular}
   \tablefoot{Model names are shortened: ``ref'' is the single-star model st31gm01n002, and ``mX'' is st31gm01n002\_mX. The columns list the primary mass $M_1$, the companion mass $M_2$, the mass ratio $q = M_2/M_1$, the tidal scale height relative to the stellar radius $h/R_\star$ (Eq.~\ref{eq:tide}), the luminosity $L$, the effective temperature $T_{\mathrm{eff}}$, the surface gravity $\log g$, the number of time snapshots $N_t$ used for the analysis, and the grid spacing $dx$.}
   \label{tab:models}
\end{table*}

\begin{figure*}
   \centering
   \includegraphics[width = \textwidth]{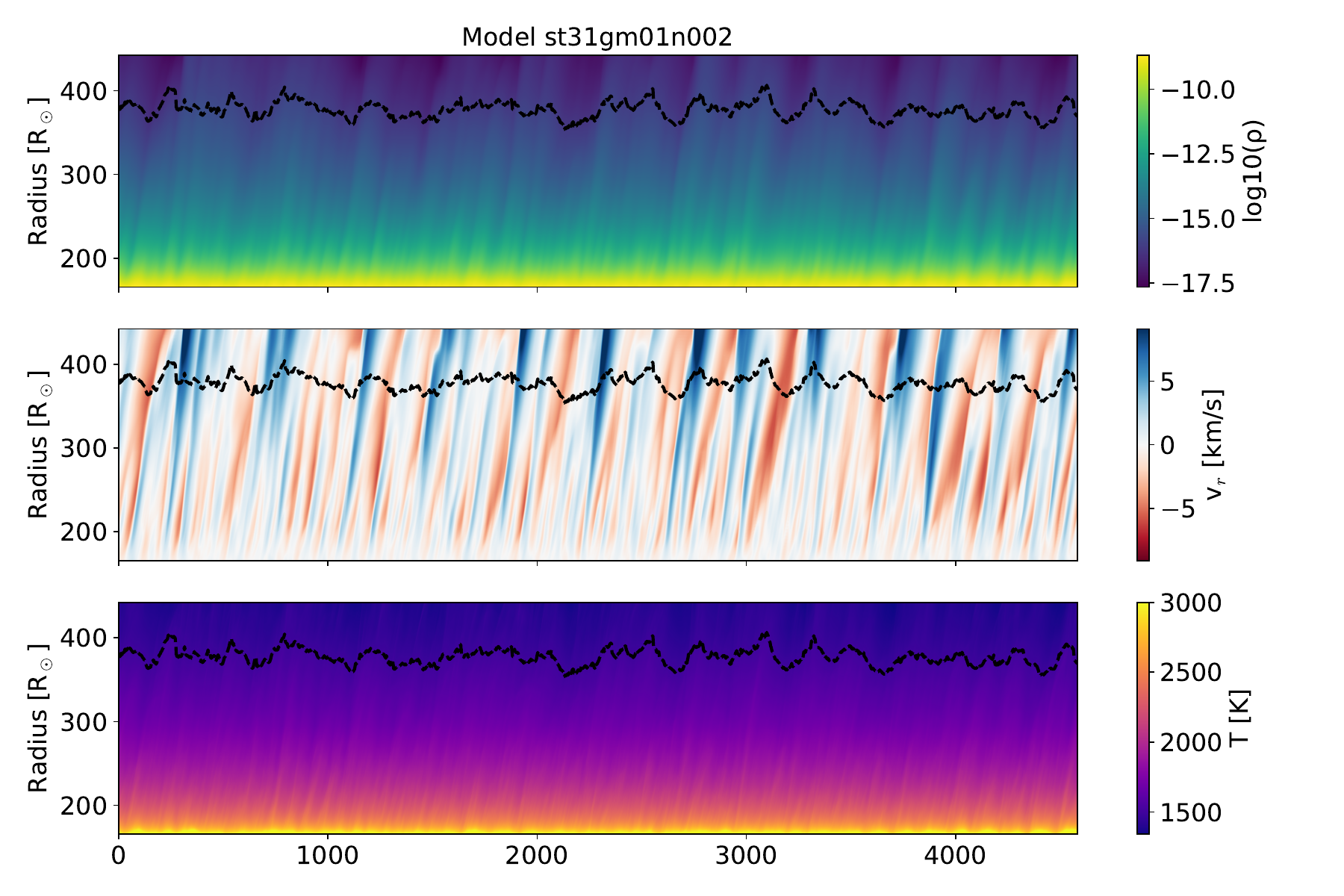}
   \caption{Spherical averages of density $\rho(r)$ (top panel), radial velocity $v_{\mathrm{rad}}(r)$ (middle panel), and Temperature $T(r)$ (bottom panel) for the reference model st31gm01n002 (no companion). The dark line corresponds to the 1500\,K isotherm, which approximate the beginning of dust condensation zone (condensation temperature of corundum).}
   \label{fig:radial_profile_nocomp}
\end{figure*}

\begin{figure*}
   \centering
   \includegraphics[width = \textwidth]{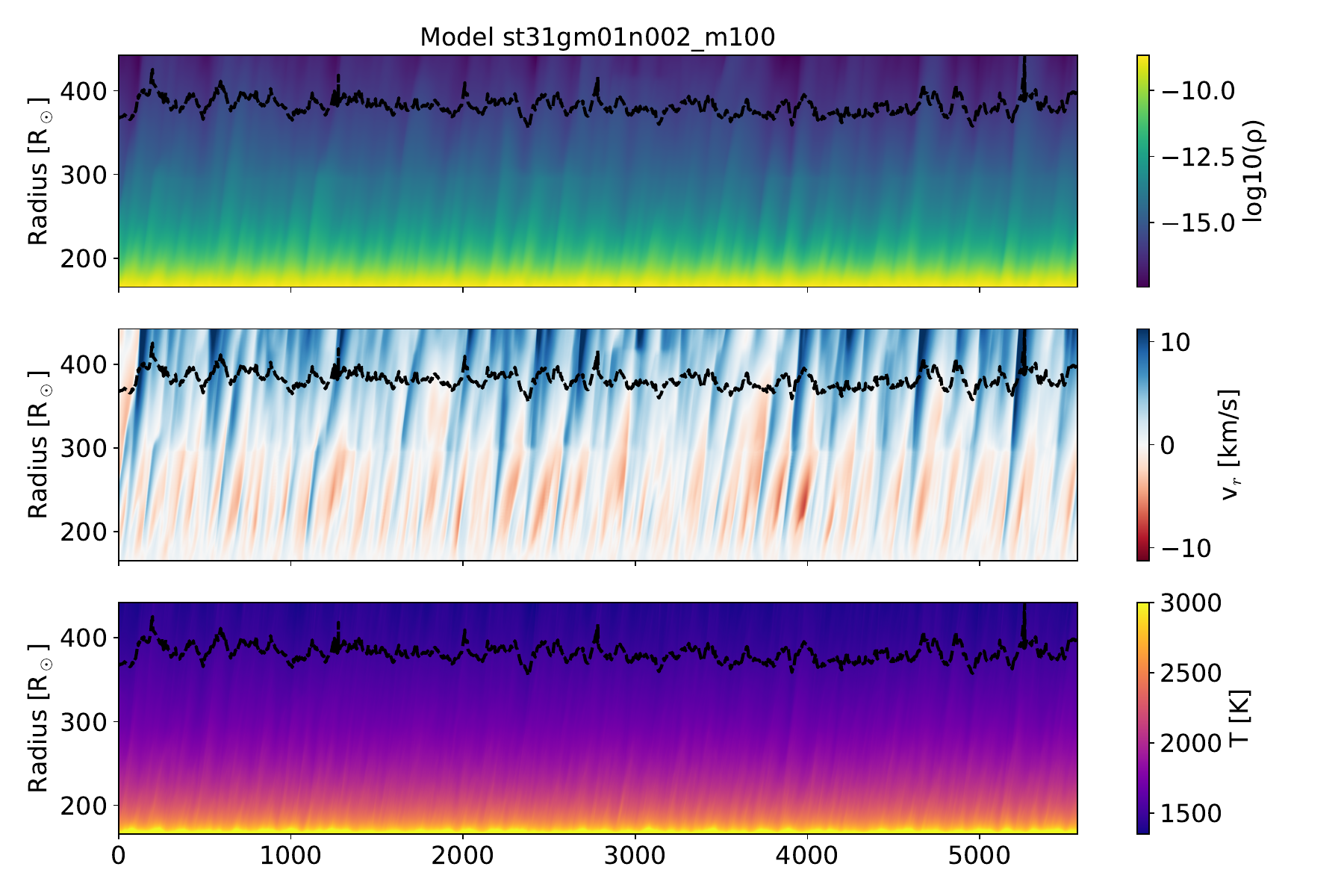}
   \caption{Spherical averages of density $\rho(r)$ (top panel), radial velocity $v_{\mathrm{rad}}(r)$ (middle panel), and Temperature $T(r)$ (bottom panel) for the most massive model st31gm01n002\_m100 (companion mass of approximately $1.7\,M_{\odot}$). The dark line corresponds to the 1500\,K isotherm, which approximate the beginning of dust condensation zone (condensation temperature of corundum).}
   \label{fig:radial_profile_m100}
\end{figure*}

\subsection{Observables and post-processing}
\label{Sec:Observables}

To convert the 3D simulation output into physically interpretable diagnostics that can be compared across the grid of binary configurations, we extract the radial velocity field $v_{\mathrm{rad}}(r,t) = \langle \rho v_{\mathrm{rad}} \rangle / \langle \rho \rangle$, derived from the radial mass flux and the density, spherically averaged over solid angle at each time step \citep{freytag2017b, Ahmad2023}, and decompose it into a set of dynamical regimes, each characterised by its pulsational properties (frequency, spectral linewidth, and spatial energy distribution). This is done in two steps: first, a Singular Value Decomposition (SVD) identifies spatially coherent dynamical modes and separates them into physically distinct regimes; second, within each regime, a Lomb--Scargle periodogram analysis is applied to extract these quantitative pulsational properties.

\subsubsection{Singular Value Decomposition (SVD) of radial velocities}
\label{Sec:Observables_SVD}

Following \cite{freytag2017b} and \cite{Ahmad2023}, we cast the spherically averaged radial velocity as a data matrix $V \in \mathbb{R}^{N_r \times N_t}$, with $N_r$ radial bins and $N_t$ time snapshots giving a 2D array visible in Fig~\ref{fig:radial_profile_nocomp} middle panel. For each model, $N_r = 445$ and $N_t$ is reported in Table~\ref{tab:models}. We then decompose it via a SVD:
\begin{equation}
   V = U \, \Sigma \, W^{\mathsf{T}} \enspace,
\end{equation}
\noindent where the columns of $U$ are orthonormal spatial (radial) eigenvectors, the columns of $W$ are the corresponding orthonormal temporal eigenvectors, and $\Sigma$ is diagonal, containing the singular values $\sigma_k$ ranked in decreasing order. 

Each mode $k$ therefore corresponds to a spatial pattern $U_{:,k}$ modulated in time by $W_{:,k}$, with an amplitude set by $\sigma_k$; the fraction of the total variance of $v_{\mathrm{rad}}(r,t)$ carried by mode $k$ scales as $\sigma_k^2 / \sum_j \sigma_j^2$. This decomposition isolates orthogonal dynamical modes without imposing a priori a radial or temporal basis, making it well suited to identify spatially distinct regimes.

Applied to the reference model (no companion), the ranked SVD spectrum\footnote{ranked by decreasing $\sigma_k$} separates into two groups of modes with distinct radial eigenvector shapes and dominant temporal frequencies: the leading modes (components 1--5) have spatial eigenvectors peaking at large radii ($r \gtrsim 1.3\,R_\star$) and are dominated by a low-frequency signal, while an intermediate group (components 6--20) has eigenvectors peaking closer to the photosphere ($0.6 \lesssim r/R_\star \lesssim 1.3$) and a higher, more coherent frequency (see also Figure~7 in \cite{Freytag2023}, and the related discussion). 

We refer to these two groups as the \emph{Pulsation Zone} and \emph{Outflow Region}, respectively (Sect.~\ref{Sec:Results}), and reconstruct the corresponding filtered radial velocity fields by summing over each group of modes, $v_{\mathrm{rad}}^{(\mathrm{group})}(r,t) = \sum_{k \in \mathrm{group}} \sigma_k \, U_{:,k} \, W_{:,k}^{\mathsf{T}}$, which are used as input to the frequency analysis described below. This filtering removes the stochastic convective background and isolates the coherent oscillatory signal in each regime, allowing us to quantify its pulsational properties.

\subsubsection{Pulsational properties from Lomb--Scargle periodograms}
\label{Sec:Observables_LS}

For each radial bin $r$ and each SVD-defined regime, we compute the power spectra with a Lomb--Scargle periodogram, which for equidistant time sampling is equivalent to a Fourier transform evaluated on an oversampled frequency grid (here five points per resolution element), and which provides the smooth spectra needed for the peak fits.

Following \cite{Ahmad2025}, the resulting periodogram is smoothed with a combination of Gaussian and Savitzky-Golay filter, respectively with a width of 5 frequency bin and a frequency cut-off of 10$^{-5}$ cycles per day to suppress high-frequency numerical noise while preserving the shape and width of the fundamental peak.

Only the fundamental mode of the smoothed PSD is considered and fitted with a Gaussian function,
\begin{equation}
   f(\nu) \propto \exp\left[-\frac{(\nu - \nu_0)^2}{2 \sigma_f^2}\right] \enspace,
\end{equation}
\noindent from which we extract the central frequency $\nu_0$ and the spectral linewidth $\sigma_f$ which quantifies the coherence of the oscillation. A narrow linewidth indicates a well-defined, long-lived periodic signal, while a broad linewidth reflects a loss of phase coherence, for instance due to stochastic convective forcing or, as discussed in Sect.~\ref{Sec:Results}, tidal perturbation by a companion. 

Finally, we define the radial power density as the fraction of the total PSD contained within the fundamental mode,
\begin{equation}
   \mathcal{P}(r) = \frac{\int_{\nu_0 - \sigma_f}^{\nu_0 + \sigma_f} P(r,\nu)\, \mathrm{d}\nu}{\int_0^{\infty} P(r,\nu)\, \mathrm{d}\nu} \enspace,
\end{equation}
\noindent where $P(r,\nu)$ is the power spectral density of the radial velocity. Evaluated over the full radial range, $\mathcal{P}(r)$ therefore measures, at each radius, the fraction of the total velocity variance carried by this specific frequency band, and its peak location identifies where each dynamical regime is preferentially confined (Figure~\ref{Fig:Domain}).

\subsection{Density anisotropy extraction}
\label{Sec:Density_method}

From the Cartesian simulation cube at each time step, the cell-centre coordinates $(x,y,z)$ are converted to spherical coordinates $r=\sqrt{x^2+y^2+z^2}$, $\phi=\mathrm{atan2}(y,x)$, and $\theta=\arccos(z/r)$. Rather than interpolating, we bin the resulting $(r,\theta,\phi)$ point cloud directly: each Cartesian cell is assigned to a spherical voxel of width $\Delta r$ in radius and $1^\circ \times 1^\circ$ in $(\theta,\phi)$, and the density in each voxel is the mean of the Cartesian cells it contains. We use $N_{\mathrm{shell}}=200$ radial shells between $R_\star$ and the box diagonal, giving $\Delta r \approx 1.75\,R_\odot$, matched to the native Cartesian grid spacing $dx$ (Table~\ref{tab:models}).

On this binned density cube, we can then construct density $\rho(r)$ in different directions by averaging inside the corresponding angular bin. To improve the statistics, we used, instead of one angular bins, angular cones of opening angles 10$^\circ$. We can therefore construct 6 different density profiles $\rho(r, \theta, \phi)$ corresponding to the 6 main directions of the cube, denoted as $\rho_{+X}(r)$, $\rho_{-X}(r)$, $\rho_{+Y}(r)$, $\rho_{-Y}(r)$, $\rho_{+Z}(r)$ and $\rho_{-Z}(r)$. We can also compute three different density profiles by averaging the two opposite directions, denoted as $\rho_X(r)$, $\rho_Y(r)$ and $\rho_Z(r)$ and defined as:
\begin{equation}
   \overline{\rho}_i(r) = \frac{\rho_{+i}(r) + \rho_{-i}(r)}{2} 
   \enspace,
\end{equation}
\noindent where $i$ can be $X$, $Y$ or $Z$.

From these density profiles, we construct two probes for asymmetry: the first one is the Binary axis / transverse density ration defined as

\begin{equation}   
   \frac{\rho_{X}}{\langle \rho_{Y,Z} \rangle }= \frac{\rho_X(r)}{\frac{1}{2}[\rho_Y(r) + \rho_Z(r)]} \enspace,
\end{equation}

\noindent Which quantifies the density contrast between the binary axis and the transverse direction. We denote $\langle \rho_{Y,Z} \rangle$ as $ \rho_{\perp} $. 

The second one is the inter-axis anisotropy defined as the ration between two different axes of the cube

\begin{equation}
   \frac{\rho_{i}}{\rho_{j}} \; \text{with } i,j = X,Y,Z \enspace,
\end{equation}

with $i$ and $j$ representing different axes of the cube ( i.e., $X$, $Y$, or $Z$). This ratio quantifies the density contrast between two axes of the cube. 

\section{Results}
\label{Sec:Results}

\subsection{Dynamical regimes of the reference AGB atmosphere}
\begin{figure}[h!]
\centering 
\includegraphics[width=0.5\textwidth]{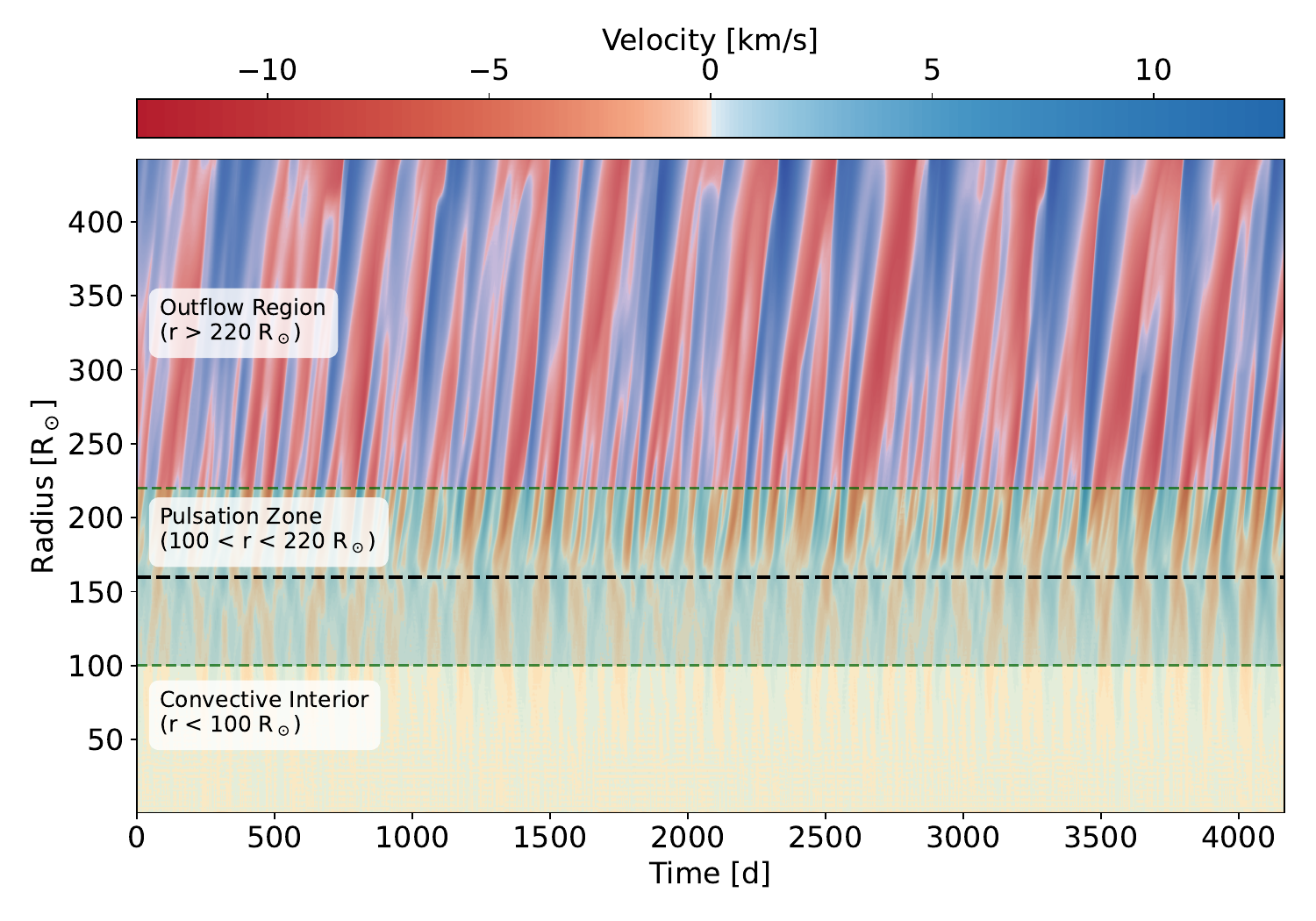} 
\caption{Radial velocity $v_{\mathrm{rad}}(r,t)$ of the reference model st31gm01n002 (no companion) filtered using SVD components.}
\label{Fig:Spatial}
\end{figure}

Before investigating binary effects, we decompose the dynamics of the single-star model. Using SVD on $v_{\mathrm{rad}}(r,t)$, we identify three spatially distinct regimes (see Figure~\ref{Fig:Spatial}):

\begin{itemize}
   \item \textbf{Convection-dominated layers ($r < 0.6\,R_\star$):} This region exhibits chaotic behaviour driven by convective motions, showing no significant periodic signal.
   \vspace{1ex}
   \item \textbf{Pulsation-dominated layers ($0.6 < r < 1.3\,R_\star$):} These layers are dominated by a clear fundamental mode with a frequency $\nu_1 \approx 0.0074\,\mathrm{day}^{-1}$ (period $P \approx 135$\,d), consistent with the findings of \cite{Ahmad2025}. In terms of SVD decomposition, this regime is captured by components 6 through 20.
   \vspace{1ex}
   \item \textbf{Outflow Region ($r > 1.3\,R_\star$):} The dynamics here are characterized by lower-frequency variability ($\nu_2 \approx 0.0039\,\mathrm{day}^{-1}$, $P \approx 256$\,d), tracing the ballistic motion of the ejected material. This region corresponds to the dominant SVD components 1 to 5.
\end{itemize}

For each model, hence for each companion mass, we compared the characteristic periods of the two pulsation peaks, $T_1=1/\nu_1$ and $T_2=1/\nu_2$, their widths $\sigma_T$, and their spatial energy distribution to assess the impact of a companion on the pulsation. Quantities related to $\nu_1$ are drawn in green and those related to $\nu_2$ in purple. The error bars are the scatter between three consecutive time segments of each run, divided by $\sqrt{3}$.

\begin{figure}[h!]
   \centering
   \includegraphics[width=0.48\textwidth]{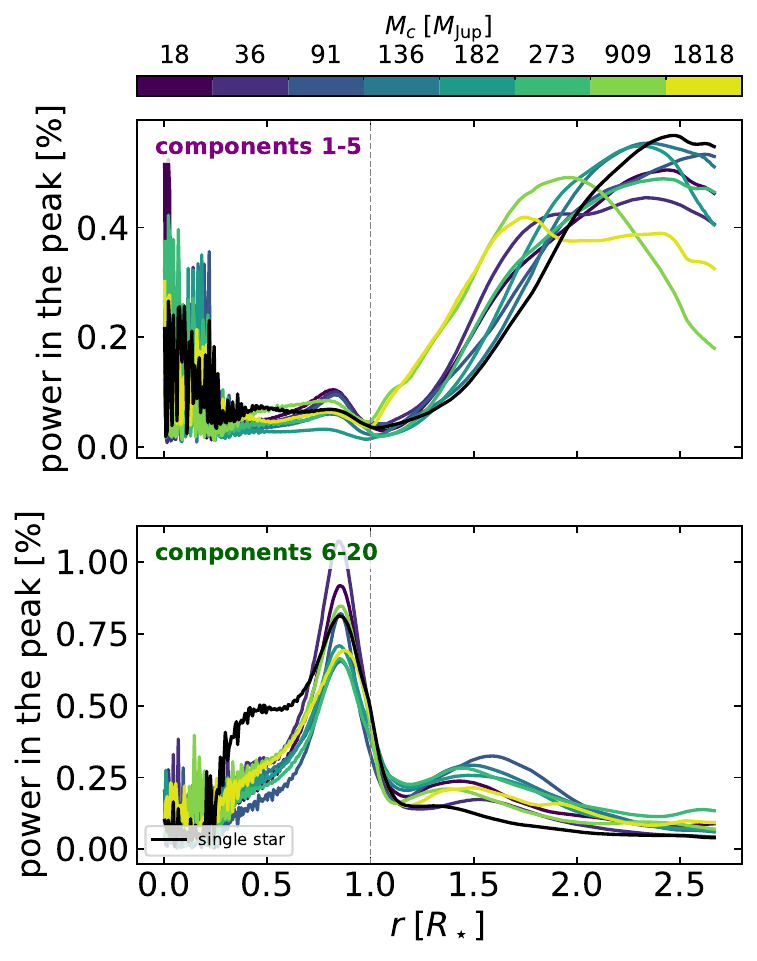}
   \caption{Fraction of the power of the peaks of $\nu_1$ (top, SVD components 6--20) and $\nu_2$ (bottom, components 1--5) as a function of radius, for all models. The colour indicates the companion mass. The single star is in black. The dashed line marks the stellar radius.}
   \label{Fig:Domain}
\end{figure}

\begin{figure}[h!]
\centering 
\includegraphics[width=0.48\textwidth]{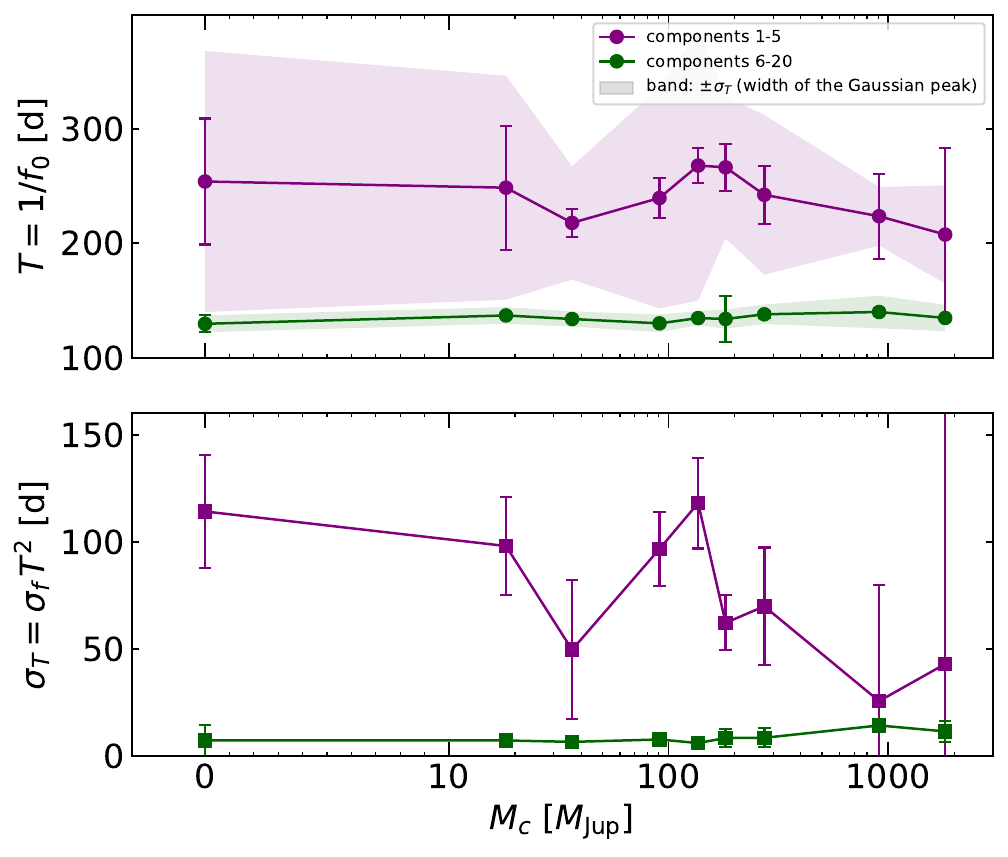} 
\caption{Characteristic period $T=1/\nu_0$ (top) and width of the peak $\sigma_T=\sigma_f\,T^2$ (bottom) as a function of companion mass. Components 6--20 ($\nu_1$) are shown in green and components 1--5 ($\nu_2$) in purple.}
\label{Fig:FreqSigma}
\end{figure}

Across all simulated models, the characteristic frequency $\nu_1 \approx 0.0074\,\mathrm{day}^{-1}$ ($T_1\approx135$\,d, captured by SVD components 6--20) remains independent of the companion mass within the uncertainties (Figure~\ref{Fig:FreqSigma}, top panel, green): the periods of all models lie between 130 and 138\,d, with errors of $\sim$20\,d, and no model differs from the single star by more than $1\sigma$. The power peak for this mode remains localized around $r\approx0.85\,R_\star$ for all companion masses (Figure~\ref{Fig:Domain}, top panel).

The same holds for the $\nu_2$ mode (purple), which is associated with the motion of the ejected material in the outer layers. Its characteristic frequency $\nu_2\approx0.0039\,\mathrm{day}^{-1}$ ($T_2\approx240$\,d, SVD components 1--5) shows no significant shift, with periods between 198 and 261\,d and errors of 10--40\,\% (up to 100\,d for the 18 and 1820\,M$_{\rm Jup}$ models), where the two largest errors reflect the short segments relative to the period. The spatial energy distribution of this mode shows some differences between models, but with no systematic dependence on the companion mass (Figure~\ref{Fig:Domain}, bottom panel); the power peak is consistently located around $r\approx2.4\,R_\star$.

The widths of the peaks (Figure~\ref{Fig:FreqSigma}, bottom panel) show minor changes for the stellar-mass companions: the peak of $\nu_2$ appears narrower and that of $\nu_1$ somewhat broader. These changes are comparable to the scatter between models of planetary mass (a standard deviation of $\sim$20\,d on $\sigma_T$ of $\nu_2$) and rely on two models, so we do not consider them significant. In summary, the presence of a companion does not significantly alter the pulsation periods nor the coherence of the modes within the uncertainties of this analysis.

\subsection{Impact of companion on density distribution}
\label{Sec:Density_results}

\begin{figure}[h!]
   \centering
   \includegraphics[width=0.48\textwidth]{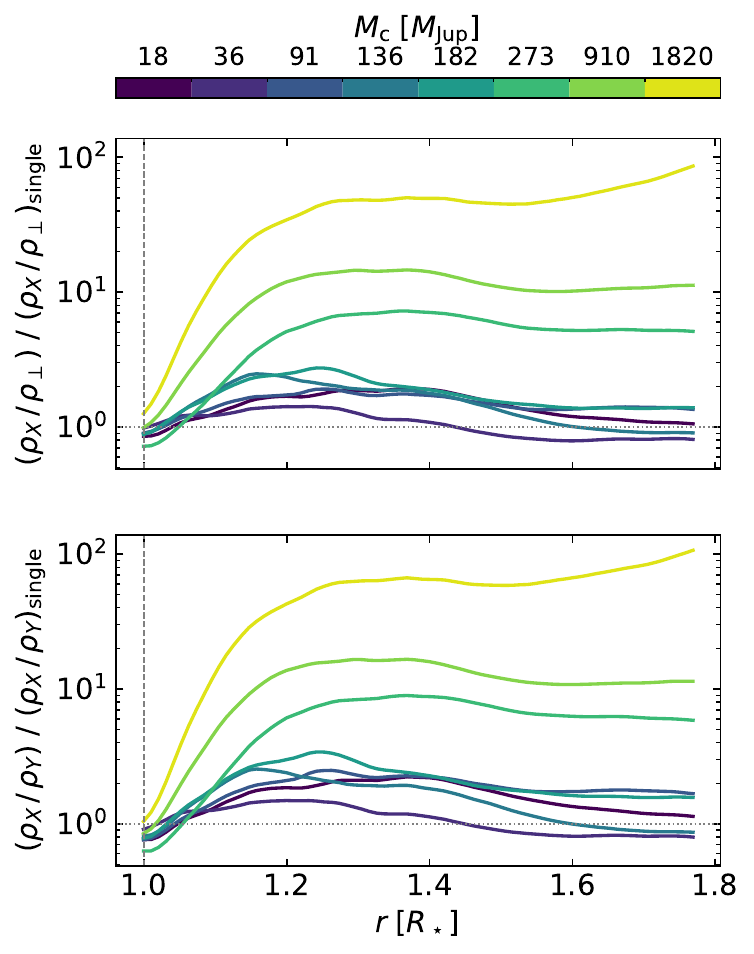}
   \caption{Density anisotropy as a function of radius for different companion masses. (top) The ratio $\rho_X / \rho_\perp$ quantifies the enhancement of density along the binary axis compared to the transverse directions. (bottom) Direct comparison between the X and Y axes.}
   \label{Fig:Density}
\end{figure}

Using the method described in Section~\ref{Sec:Density_method}, we extract the density distribution in different directions of the cube and quantify the anisotropy of the extended atmosphere. We observe that the presence of a companion systematically enhances the density along the binary axis (X-axis) compared to the transverse directions (Y and Z axes) (Figure~\ref{Fig:Density}, top panel) and inside the orbital plane (Figure~\ref{Fig:Density}, bottom panel). This anisotropy becomes more pronounced with increasing companion mass, indicating that the gravitational influence of the companion shapes the morphology of the extended atmosphere. The density contrast between the binary axis and the transverse directions reaches a factor of 5.5 for the most massive companion, and in extreme cases (e.g., $M_c = 100\,M_0 \simeq 1800$ \mjup) the density along the binary axis is up to two orders of magnitude higher than in the transverse directions.

\begin{figure}[h!]
   \centering
   \includegraphics[width=0.48\textwidth]{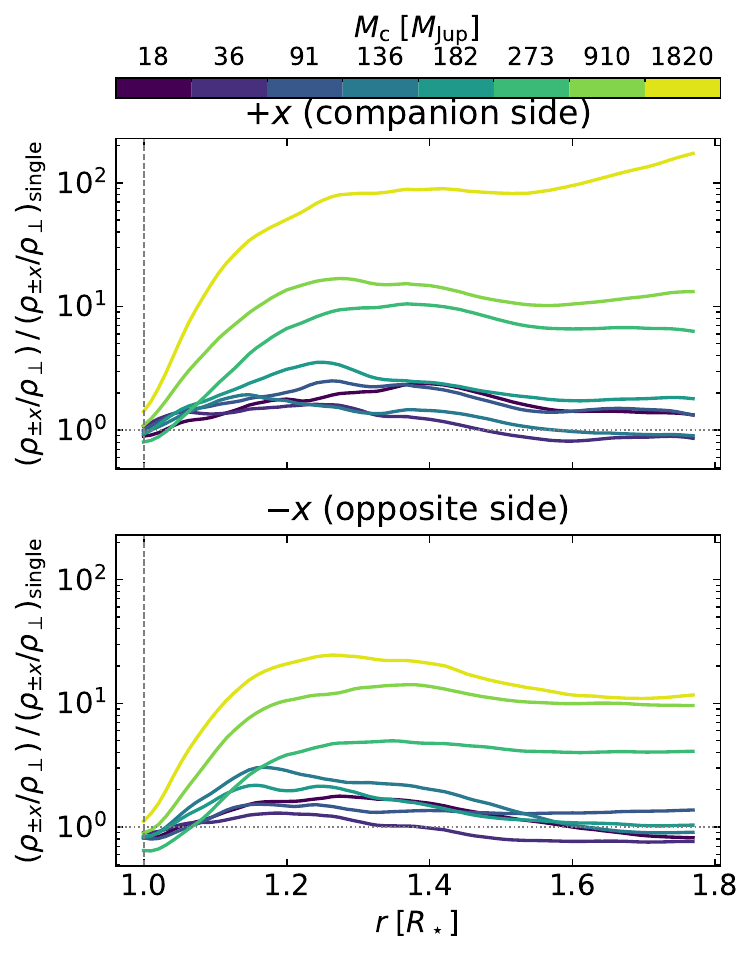}
   \caption{Density along the binary axis, on the side of the companion ($+X$, left) and on the opposite side ($-X$, right), normalised to the mean density in the four transverse directions, as a function of radius and for different companion masses. The dashed vertical line marks the stellar radius.}
   \label{Fig:TidePair}
\end{figure}

To separate the two sides of the binary axis, we also compare the density along $+X$ (towards the companion) and $-X$ (opposite side) separately to the mean density in the four transverse directions, $\rho_\perp$ (Figure~\ref{Fig:TidePair}). Both $\rho_{+x}/\rho_\perp$ and $\rho_{-x}/\rho_\perp$ increase above unity with increasing companion mass, so the companion produces two overdensities, one on each side of the star: a tide on the side of the companion and an ''anti-tide'' on the opposite side. For low-mass companions the two profiles are nearly identical. For the most massive companions the overdensity on the companion side is larger than on the opposite side, in line with the stronger gravitational pull of the companion at short distance: on the far side, the centrifugal force partly compensates the companion's gravity, and the two forces have a different dependence on radius.

Close to the stellar surface (i.e., $r < 1.1 R_\star$), the anisotropy is less pronounced, indicating that the convective motions dominate the dynamics in this region. However, as the material moves outward and the influence of convection diminishes, the gravitational effect of the companion becomes more significant, leading to a more pronounced anisotropy of the extended atmosphere.

In most cases, for both ratios, the anisotropy reaches a maximum at a radius of about $r \approx 1.2 - 1.4 R_\star$. It then reaches a plateau and decreases slowly with radius, but remains significant even at larger distances. This suggests that the companion's influence extends well beyond the immediate vicinity of the AGB star.

\begin{figure}[h!]
   \centering
   \includegraphics[width=0.48\textwidth]{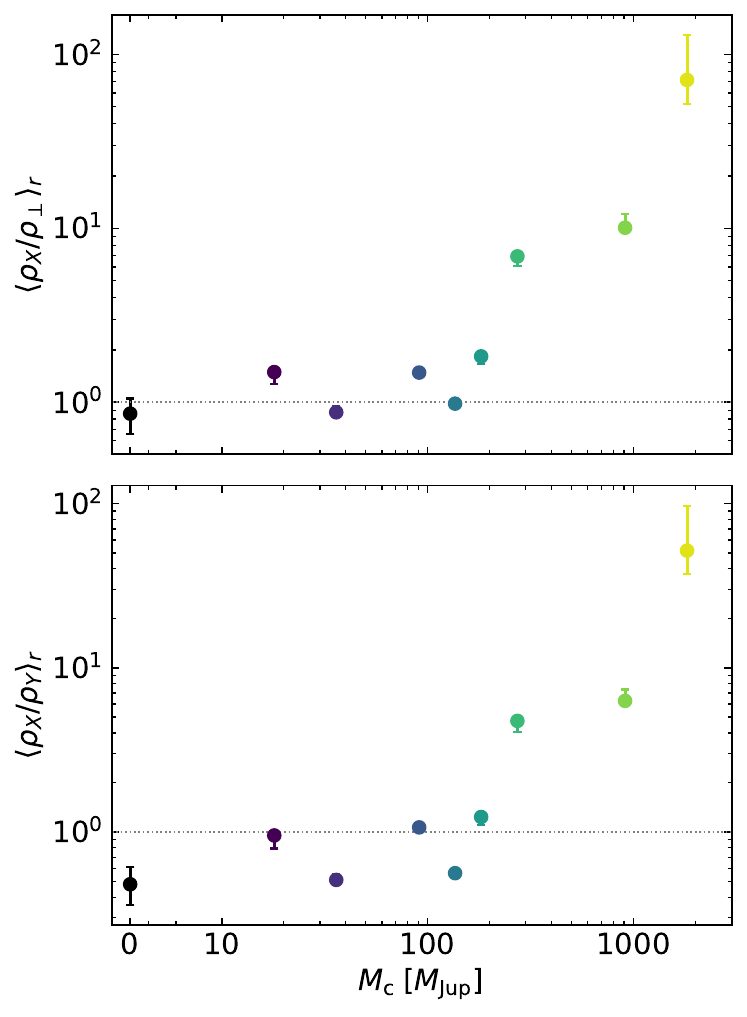}
   \caption{Mean value of each anisotropy measure as a function of companion mass. (top) The mean ratio $\rho_X / \langle \rho_{Y,Z} \rangle$ quantifies the enhancement of density along the binary axis compared to the transverse directions. (bottom) Direct comparison between the X and Y axes.}
   \label{Fig:MeanAnisotropy}
\end{figure}

By averaging the values beyond this maximum, we obtain an estimate of the mean anisotropy, shown in Figure~\ref{Fig:MeanAnisotropy} as a function of companion mass. The anisotropy increases with companion mass, confirming that the gravitational influence of the companion shapes the extended atmosphere. For the most massive companion, the mean anisotropy reaches a factor of 3 in the XY plane and up to 5 between the X-axis and the transverse directions. This behaviour suggests a preferred direction along the binary axis, which becomes more focused on the X-axis as the companion mass increases.

Our models do not include dust formation, and the star is probably too compact and hot to drive a wind without it. The enhanced density along the binary axis therefore describes material levitated in the outer atmosphere, and it does not demonstrate an enhanced mass loss in this direction. Whether this overdensity leads to a preferential outflow, and ultimately influences the shaping of the circumstellar environment, can only be tested with models that include dust and non-grey radiative transfer, which is a natural next step of this work.

\section{Discussion}

\subsection{On the L$_2$~Pup benchmark}
\label{Sec:L2Pup_caveat}

It is important to stress how closely our reference stellar model actually matches L$_2$ Pup A. The dynamical mass measured from ALMA disk kinematics, $M_A = 0.659 \pm 0.011\,(\mathrm{stat}) \pm 0.041\,(\mathrm{sys})\,M_\odot$ \citep{Kervella2016}, is $\sim8\sigma$ away from the $1\,M_\odot$ primary used throughout this study: the two are not formally consistent. This mismatch, however, should be read against the historically large spread of indirect mass estimates for L$_2$ Pup A prior to this dynamical measurement, ranging from $0.5\pm0.2\,M_\odot$ \citep{Ohnaka2014} to $0.7\,M_\odot$ \citep{Lykou2015a}, $1.7\,M_\odot$ \citep{DummSchild1998}, and $2\,M_\odot$ \citep{Kervella2014}. Our $1\,M_\odot$ reference model therefore falls within the broader historical range for this star, even though it does not match the tightest current dynamical constraint. We nonetheless adopt it because it is the most extensively tested and numerically stable single-star \cob\ configuration available that is close enough to L$_2$ Pup A configuration, already validated against AGB observables in previous work \citep{Freytag2023,Beguin2024}, rather than a dedicated, purpose-built match to L$_2$ Pup A. L$_2$ Pup should therefore be understood as providing the companion mass and separation scale for our grid (Sect.~\ref{Sec:Methods}), not as a quantitative reproduction of the system as a whole.

\subsection{Center of mass analysis}
\label{Sec:COM}
We quantify the global atmospheric displacement by tracking the center of mass $\mathbf{R}_{\mathrm{c}}$:
\begin{equation}
    \mathbf{R}_{\mathrm{c}} = \frac{\sum_{i,j,k} \rho_{i,j,k} \, \mathbf{r}_{i,j,k}}{\sum_{i,j,k} \rho_{i,j,k}} \enspace,
\end{equation}

\noindent where $\rho_{i,j,k}$ is the density in the Cartesian cell at $(i,j,k)$ and $\mathbf{r}_{i,j,k}$ is its position vector. The sum runs over all cells in the computational domain, to capture the global displacement of the atmosphere and not only the local density enhancement along the binary axis. The center of mass is computed every 40 time steps (i.e., every 0.92 days).

Following the methodology of \cite{Beguin2024}, but applying it to the center of mass rather than the photocenter, we calculate the temporal mean offset and standard deviation of $\mathbf{R}_{\mathrm{c}}$ along each axis. The uncertainties are obtained with a block bootstrap that accounts for the temporal correlation of the series, and we checked their reliability against the variation of the standard deviation between consecutive segments of the same run.

The tidal deformation of the star is symmetric with respect to the stellar centre: the star is slightly elongated along the axis towards the companion, on both sides. It should therefore not displace the center of mass, but it could increase the fluctuations of $\mathbf{R}_{\mathrm{c}}$ along the binary axis. For our most massive companion ($1.7\,M_\star$), the expected tidal height is $h \approx 3\,\%\,R_\star$.

\begin{figure}[h]
   \centering
   \includegraphics[width=0.45\textwidth]{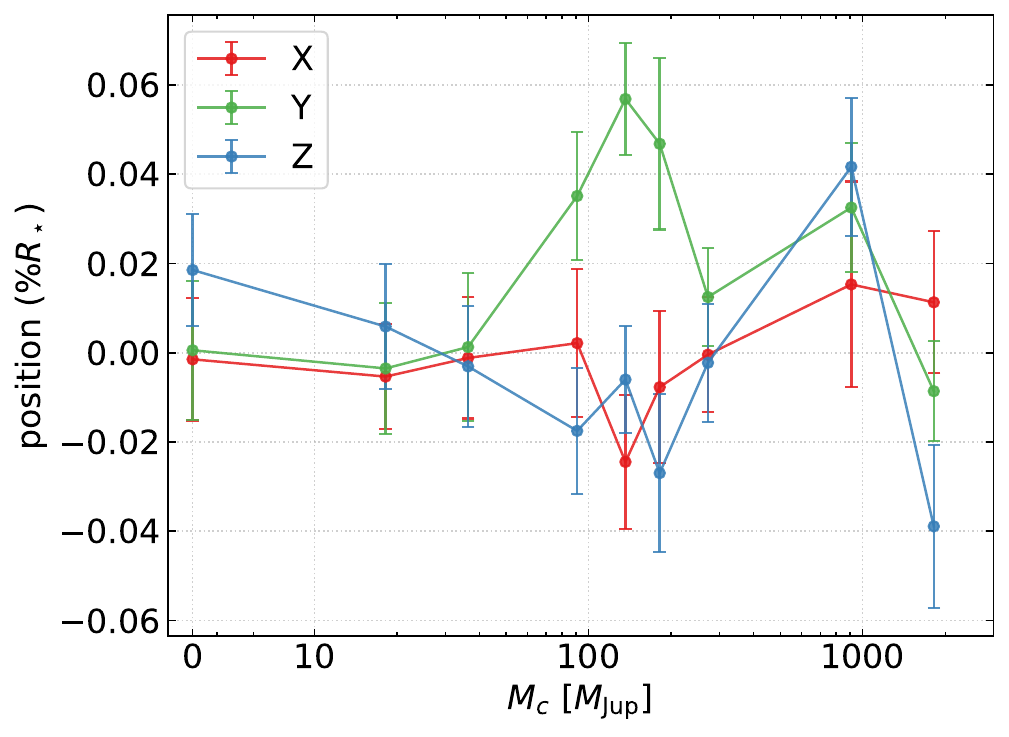}
   \caption{Mean position of the center of mass $\mathbf{R}_{\mathrm{c}}$ as a function of companion mass $M_c$.}
   \label{Fig:COM}
\end{figure}

\begin{figure}[h]
   \centering
   \includegraphics[width=0.48\textwidth]{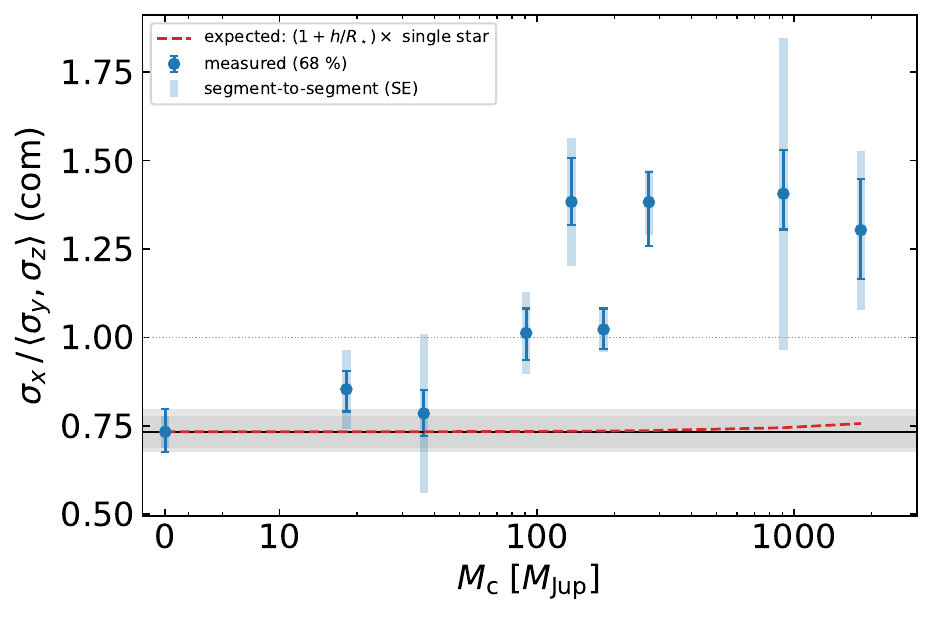}
   \caption{Anisotropy of the center of mass fluctuations, defined as the ratio of the standard deviation along the binary axis to the mean of the standard deviations along the two transverse axes, as a function of companion mass $M_c$.}
   \label{Fig:COM_aniso}
\end{figure}

In Figure~\ref{Fig:COM} we display the mean position of the center of mass and its standard deviation for different companion masses. Neither the mean offset nor the standard deviation of $\mathbf{R}_{\mathrm{c}}$ exhibits a discernible trend with the companion mass. Throughout our grid, the mean offset remains small, at approximately $0.05\,\%\,R_\star$, while the fluctuations (standard deviation) are of the order of $2\,\%\,R_\star$. The standard deviation along the binary axis does not grow compared to the transverse directions beyond the variations seen between independent segments of the same run (Figure~\ref{Fig:COM_aniso}).

The fluctuations of $\mathbf{R}_{\mathrm{c}}$ are produced by the large-scale convective flows. The standard deviation varies by 30--50\,\% between consecutive segments of a single run, well above the 3\,\% expected for the tidal elongation. The center of mass is therefore not a sensitive probe of the companion in these simulations, and we do not attempt to infer its effect from it.

All models with a companion include the centrifugal term of the modified potential (synchronous rotation at the orbital frequency), which is absent in the single-star reference. This term is independent of the companion mass, and it acts on the stellar surface differently on the two sides of the star (larger away from the companion), which partly compensates the difference in the gravitational pull of the companion. Its effect is small, and the lowest-mass model ($18\,M_{\rm Jup}$), where the tide is negligible, serves as a benchmark for the effect it would have on the single star model.

\subsection{Non-radial modes in the stellar-mass test cases}
\label{Sec:LMmodes}
\begin{figure}
   \includegraphics[width=0.45\textwidth]{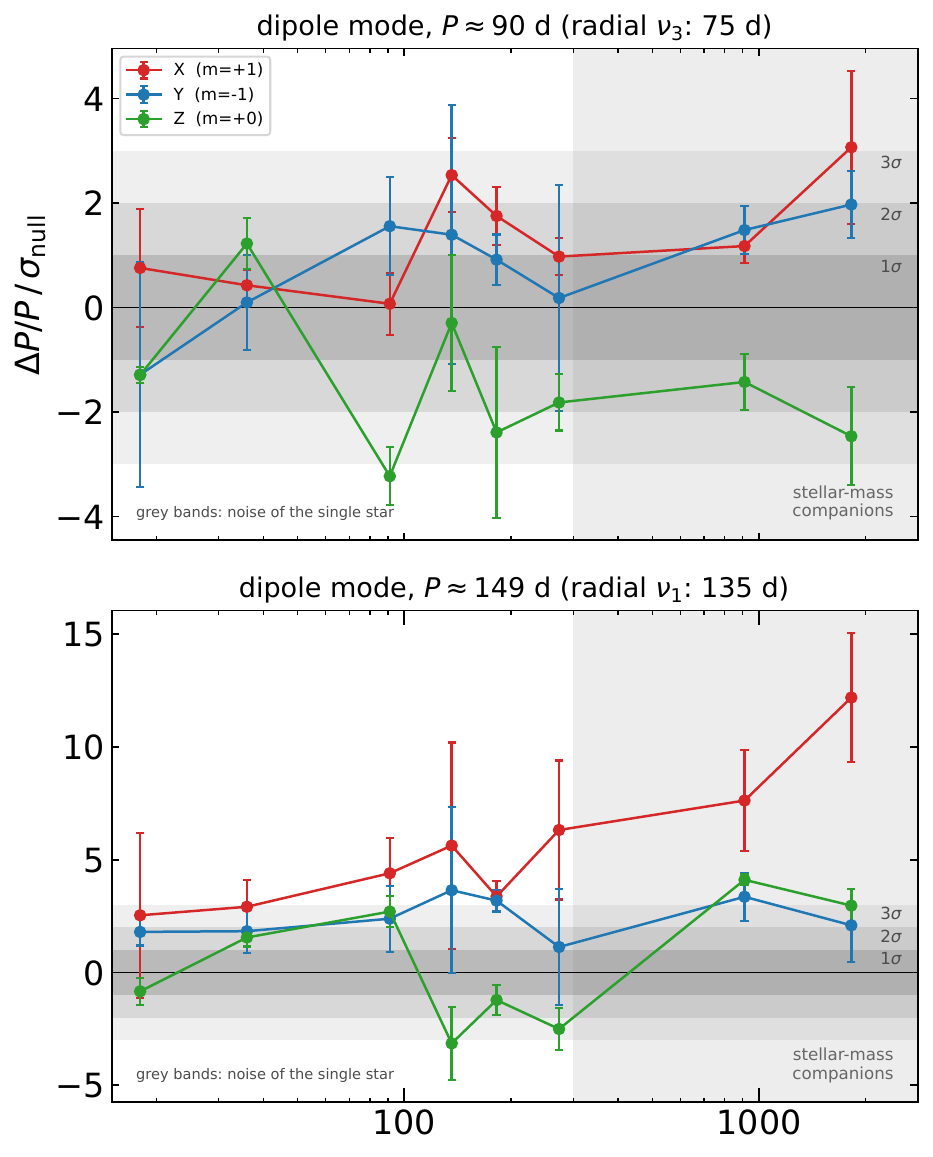}
   \caption{Period shift of the dipole ($l{=}1$) pulsation modes with respect to the single-star model, along the binary axis ($X$, $m{=}+1$) and the two perpendicular axes ($Y$, $m{=}-1$; $Z$, $m{=}0$), as a function of companion mass, for the two main pulsation peaks. The shift is divided by the noise of the single star ($\sigma_{\rm null}$); the grey bands mark 1, 2, and 3$\sigma$. Error bars are the scatter between independent time segments. The shaded region indicates the stellar-mass companions.}
   \label{fig:l1_shift}
\end{figure}

We extended the period analysis to the non-radial modes. For each model and each $(l,m)$ component, we measured the shift of the pulsation peaks relative to the single-star model, using the cross-correlation of the power spectra, and expressed it in units of the noise of the single star, $\sigma_{\rm null}$. This noise is estimated by comparing independent time segments of the single-star run with one another. Figure~\ref{fig:l1_shift} shows this normalised shift for the dipole mode ($l{=}1$) along the three axes ($m{=}+1$, $-1$, $0$ i.e. $X$, $Y$, $Z$), for each of the two main pulsation peaks. The $X$ component, aligned with the binary axis, lies furthest above the noise, mostly for the second peak, while $Y$ and $Z$ stay closer to the single-star values. This trend should be taken with caution. No component shows a significant slope for companion masses up to $\sim$270\,M$_{\rm Jup}$, the trend relies on the two stellar-mass models, and a smaller shift is already present at 18\,M$_{\rm Jup}$, where the tide is negligible, so it reflects the model-to-model scatter rather than the companion. Longer simulations and more models are needed to confirm that the aligned mode is affected by the tide.

\subsection{Luminosity Anisotropy}
\label{Sec:LumAniso}
To test whether the companion leaves a signature in the emergent flux, we computed the bolometric luminosity leaving each face of the computational box and compared the flux through the two faces perpendicular to the binary axis to that through the four transverse faces:

\begin{equation}
    \frac{L_X}{L_\perp} = \frac{L_X}{\langle L_Y, L_Z \rangle} \enspace,
\end{equation}

\noindent where $L_i = \langle L_{+i}, L_{-i} \rangle$ with $i = X,Y,Z$. 

\begin{figure}[h]
    \centering
    \includegraphics[width=\linewidth]{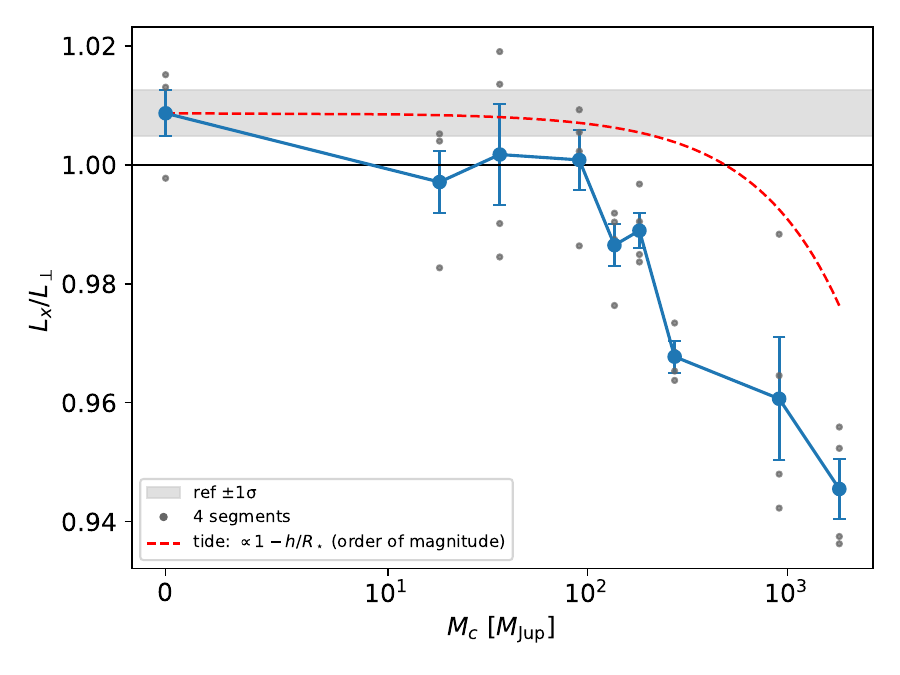}
    \caption{Ratio of the flux through the two faces perpendicular to the binary axis to that through the transverse faces, $L_X/\langle L_Y,L_Z\rangle$, as a function of companion mass $M_c$. Grey dots: values measured in independent time segments; error bars: noise from the scatter between segments. The red dashed line shows the order of magnitude of the decrease expected from the tidal elongation of the star alone.}
    \label{fig:LumAniso}
\end{figure}

The total luminosity ($\sim$2500\,$L_\odot$) does not change with the companion mass (variation of $+0.45\pm0.26$\,\% for the most massive companion), and we find no significant difference between the $+x$ and $-x$ faces. In contrast, the ratio $L_x/L_\perp$ decreases with the companion mass (Figure~\ref{fig:LumAniso}). Relative to the single star, it is unchanged within the noise for companions up to 91\,M$_{\rm Jup}$ (changes of $-0.7$ to $-1.2$\,\%, to be compared with a scatter of $\sim$0.8\,\% between independent segments of the single-star run), and then drops by 2.0--2.2\,\% for 136--182\,M$_{\rm Jup}$, by 4.1\,\% for 273\,M$_{\rm Jup}$, and up to 6.3\,\% for the most massive companion. This is 4 to 10 times the estimated uncertainty (Appendix~\ref{app:errors}).

The decrease is symmetric in $\pm x$, as expected for a tidal (quadrupolar) effect, but larger than the tidal elongation of the star alone would give (0.2--3\,\% for the same masses). We attribute this excess to the denser material above the surface (Sect.~\ref{Sec:Density_method}, Fig.~\ref{Fig:TidePair}) rather than to the deformation of the surface. The opacity depends non-linearly on temperature and density in the transition between the regime dominated by atomic lines (``clear sky'') and the one dominated by molecular lines (``thick clouds''), so a moderate density increase along $\pm x$ can reduce the emergent flux more than proportionally. This is the same type of effect as the dimming events found in 3D simulations of red supergiants by \citet{Freytag2024}, who suggested a link with the brightness variations of AGB stars. Since the time-averaged luminosity is unchanged across all models (within 0.5\,\%), the lower flux through the faces perpendicular to the binary axis is compensated by a higher flux through the transverse faces. The flux is then redistributed between directions rather than lost. This differs from a dimming event, where the spherically averaged luminosity drops. This is the clearest signature of the companion in our simulations, and it appears in an observable quantity, the bolometric intensity.
Because the effect is symmetric in $\pm x$, it would modulate the brightness twice per orbit. Some long secondary periods (LSPs) seen in pulsating red giants, often attributed to dust clouds orbiting with a companion, could therefore be explained by a change of surface brightness instead. This could be tested with existing lightcurve databases, preferably in optical bands where the molecular opacity effect should be strongest. We stress that this is an interpretation: our simulations are performed in the co-rotating frame, so the box faces do not represent the view of a distant observer, and a proper test requires synthetic observations of the models.

\subsection{Mass loss and Dust Formation}

Our simulations are gas-only and do not include dust formation, so we cannot assess how a companion affects dust condensation or the resulting mass-loss rate. A dust-driven wind, as implemented in \cite{Freytag2023}, requires non-grey radiative transfer and a treatment of the dust that are beyond the scope of this first investigation. We can nevertheless speculate on what our results imply for it. Dust formation depends most sensitively on temperature, but once the temperature is favourable, the gas density sets the time scale of the condensation. A star that forms dust only sporadically and in small amounts, given its mass and effective temperature, could therefore form it more efficiently towards, and away from, a sufficiently massive and close companion.

At $r\approx2\,R_\star$ the gas density relative to the single-star model rises by a factor of $\sim$4 on the companion side and $\sim$3 on the opposite side for the 273\,M$_{\rm Jup}$ companion, and by $\sim$12 and $\sim$10 for 910\,M$_{\rm Jup}$ (Figure~\ref{Fig:TidePair}). For 1820\,M$_{\rm Jup}$ it reaches $\sim$100 on the companion side and $\sim$8 on the opposite side. Below $\sim$200\,M$_{\rm Jup}$ the changes remain within the scatter of the single-star model. At the same time, the density in the transverse directions drops to about half of its single-star value for the most massive companions, so the gas is redistributed towards the binary axis. Whether an increase of this size changes the dust formation significantly cannot be decided with gas-only models. If it does, the two overdensities would produce two episodes of enhanced dust formation per orbit, on opposite sides of the star, rather than the single one expected for a process directly related to the companion.

This is relevant for observed AGB stars with close companions. For $\pi^1$~Gru, \cite{Drevon2026} image dust and mass transfer towards the companion with VLTI/MATISSE and favour a wind Roche-lobe overflow scenario. The system ($M_{\rm A}=1.12\,M_\odot$, $R_{\rm A}=334\,R_\odot$, $M_{\rm C}=1.18\,M_\odot$, separation 6.81\,AU) has a separation of $4.4\,R_{\rm A}$, and the AGB star fills about 60\,\% of its Roche lobe. This is close to the configuration of our stellar-mass companions (separation $4.3\,R_\star$, 59\,\% for the 910\,M$_{\rm Jup}$ companion). Wind Roche-lobe overflow requires dust formation in the extended atmosphere to accelerate the wind within the Roche lobe of the companion. The density increase along the binary axis found in our models would favour dust formation exactly where such a transfer takes place, and we therefore expect the effect of a companion on the mass loss to appear first through dust formation. This should be tested with models that include dust.

\subsection{Numerical limitations}
\label{Sec:Limitations}

Throughout this study we use several independent diagnostics to trace the atmospheric dynamics and structure response to the introduction of a companion. 
Of these tracers, only the density distribution behaves this way: the density contrast between the binary axis and the transverse directions increases systematically as the companion mass increases (Sect.~\ref{Sec:Density_method}). The pulsation properties of the star show no such behaviour -- neither the fundamental frequency nor its spectral linewidth grow with companion mass -- and neither does the center-of-mass analysis (Sect.~\ref{Sec:COM}): neither the mean offset nor its scatter increase as the companion becomes more massive, indicating that the global displacement of the atmosphere is not measurably affected by the companion over the mass and separation range explored here. 

A key physical effects is also neglected in our modified potential: the Coriolis force. Indeed, it is absent from the equations of motion, even though the centrifugal term is included. The Coriolis force imparts tangential momentum to the gas as it flows outward, causing it to spiral around the centre of mass rather than move radially. This effect could reshape or wind up the density and velocity asymmetries reported here. Neglecting it likely biases our results toward purely radial, non-rotating outflow morphologies. 

The size of the computational domain is a further limitation. We restrict our models to a $3.5\,R_\star$ box, sufficient to capture the convective interior and the pulsation driven atmosphere, but too small to follow the outflow to the point where it becomes fully unbound, or to capture its interaction with the companion, which orbits just outside the domain at $d\approx4.3\,R_\star$. Extending the computational domain, as in \cite{Freytag2023}, is therefore a priority for future work, together with the inclusion of the Coriolis force.

\section{Conclusions}

Using \cob\ with a binary-adapted potential, we produced simulations motivated by the parameters of the L$_2$ Pup system, and investigated how a companion affects the convective interior and the atmosphere of an AGB star across a grid of companion masses ($18$--$1820\,\mjup$, fixed separation of $4.3\,R_\star$). The models are gas-only and do not include dust formation. Our main findings are:

\begin{itemize}
   \item The companion produces two overdensities along the binary axis, a tide on the companion side and an ``anti-tide'' on the opposite side. The density anisotropy increases monotonically with companion mass, up to a factor of $\sim$5 between the binary axis and the transverse directions (Sect.~\ref{Sec:Density_results}). At $r\approx2\,R_\star$ the density along the axis exceeds that of the single-star model by factors of $\sim$4-12 for the 273 and 910\,$\mjup$ companions, and by $\sim$100 on the companion side for 1820\,$\mjup$, while it remains within the scatter of the single-star model below $\sim$200\,$\mjup$. This is the most robust signature of binarity found in this study.
   \vspace{1ex}
   \item The bolometric luminosity of the star is unchanged, but the flux through the faces of the box perpendicular to the binary axis decreases relative to the transverse faces, by 2.5--6.4\,\% for companions of $136$--$1820\,\mjup$ (Sect.~\ref{Sec:LumAniso}). The effect is symmetric in $\pm x$ and larger than the tidal elongation of the star alone, which we attribute to the non-linear dependence of the opacity on temperature and density in the denser atmosphere along the binary axis. This is the clearest signal in an observable, and it can in principle be improved by averaging over orbital phases and by the choice of the filter band.
   \vspace{1ex}
   \item Using SVD, we identify three dynamical regimes in the AGB atmosphere and its surroundings: convection-dominated layers ($r < 0.6\,R_\star$), pulsation-dominated layers ($0.6 < r < 1.3\,R_\star$) dominated by a fundamental mode at $\nu_1 \approx 0.0074\,\mathrm{day}^{-1}$, consistent with the fundamental mode found by \cite{Ahmad2025}, and an outflow region ($r > 1.3\,R_\star$) dominated by a lower-frequency mode at $\nu_2 \approx 0.0039\,\mathrm{day}^{-1}$ (Sect.~\ref{Sec:Observables}). The periods of the two peaks do not change significantly with companion mass for companions up to $\sim$270\,$\mjup$. For the stellar-mass test cases, the dipole mode aligned with the binary axis shows a small shift towards longer periods, but the signal is weak compared to the noise and relies on only two models (Sect.~\ref{Sec:LMmodes}).
   \vspace{1ex}
\end{itemize}

Since our models do not include dust, we cannot test how the overdensity along the binary axis affects the mass loss. Dust formation depends most sensitively on temperature, but once the temperature is favourable, the gas density sets the time scale of the condensation, so a density increase of the size we find may make dust formation more efficient towards and away from a sufficiently massive and close companion. This is relevant for wind Roche-lobe overflow in systems such as $\pi^1$~Gru \citep{Drevon2026}, whose separation in units of the AGB radius is close to that of our models. These results motivate three priorities for future work: (i) a control model with the centrifugal term but no companion, and simulations including the Coriolis force, to separate tidal effects from the effect of the modified potential; (ii) a larger computational domain, to follow the outflow past the companion's orbit; and (iii) include the dust in our binary models as in \cite{Freytag2023}, with non-grey radiative transfer, to test whether a companion affects the dust-formation zone and the mass-loss rate.

L$_2$ Pup and $\pi^1$~Gru remain the only AGB stars with a dynamically measured companion, but the Gaia DR3 non-single-star catalogue \citep{GaiaNSS2023} offers a path to further benchmarks, provided future models cover companion masses and separations for which the effect of the companion on the atmosphere is strong enough to be measured.

\bibliographystyle{aa}
\bibliography{bibliography}

\begin{acknowledgements}
B.F. acknowledges support from the Knut and Alice Wallenberg Foundation under grant no.~KAW~2023.0314. Some computations were enabled by resources provided by the Swedish National Infrastructure for Computing (SNIC) and the National Academic Infrastructure for Supercomputing in Sweden (NAISS), partially funded by the Swedish Research Council through grant agreement no.~2022-06725. This work was granted access to the HPC resources of Observatoire de la Côte d’Azur, Mésocentre SIGAMM. This work is supported by the French National Research Agency (ANR) funded project PEPPER (ANR-20-CE31-0002. 
\end{acknowledgements}

\begin{appendix}
\FloatBarrier
\nolinenumbers
\section{Tidal height derivation}
\label{Sec:tide}
We consider the primary AGB star with mass $M_\star$ and radius $R_\star$, and a companion with mass $M_c$ located at a distance $d$. We define a reference frame centred on the primary star. The gravitational potential exerted by the companion at a position $\mathbf{r}$ near the surface of the primary (where $r = |\mathbf{r}| \ll d$) can be expanded in Legendre polynomials.

The total potential includes the gravitational pull of the companion and the non-inertial centrifugal term due to the orbital motion of the primary around the center of mass. The zeroth-order term is constant and does not affect the shape. The first-order term corresponds to the uniform acceleration of the primary towards the companion and cancels out in the co-moving frame. The leading term responsible for the deformation is the second-order tidal potential $\Phi_{\mathrm{tide}}$:

\begin{equation}
   \Phi_{\mathrm{tide}}(\mathbf{r}) = -\frac{G M_c}{d^3} r^2 P_2(\cos \theta) = -\frac{G M_c r^2}{2 d^3} (3 \cos^2 \theta - 1) \enspace,
\end{equation}

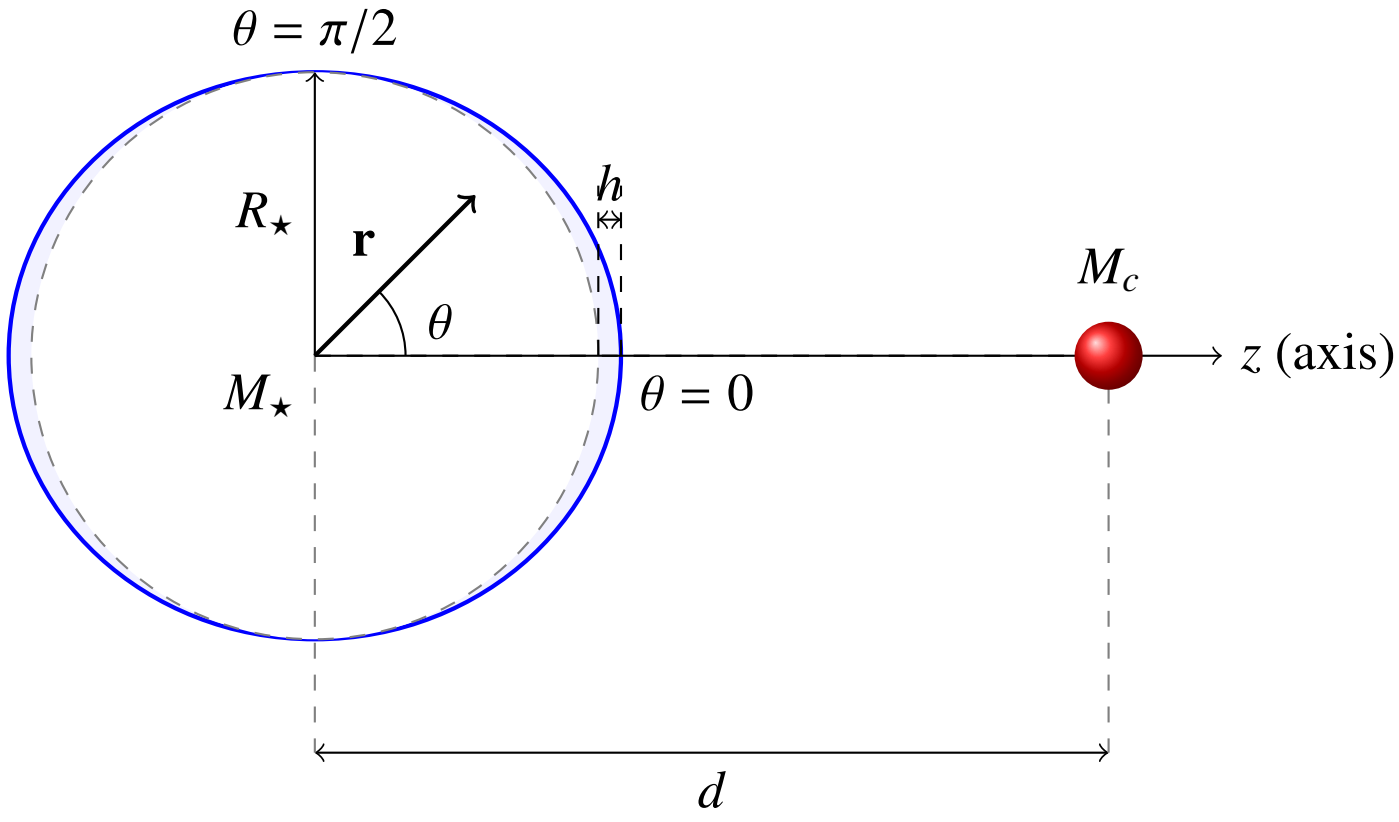
\begin{figure}[h]
   \centering
   \begin{tikzpicture}[scale=0.75]
    \def\R{2.5} 
    \def\d{7} 
    \def\h{0.2} 

    \coordinate (O) at (0,0);
    \coordinate (C) at (\d,0);

    \draw[thick, blue, fill=blue!5] (O) ellipse ({\R+\h} and \R);
    
    \draw[dashed, gray, fill=white!20] (O) circle (\R);
    \node at (O) [below left] {$M_\star$};
    
    \draw[<->] (0,-3.5) -- (\d,-3.5) node[midway, below] {$d$};
    \draw[dashed, gray] (0,0) -- (0,-3.5);
    \draw[dashed, gray] (\d,0) -- (\d,-3.5);

    \draw[->] (O) -- (0,\R) node[midway, left] {$R_\star$};
    \node at (0, \R) [above] {$\theta=\pi/2$};

    \draw[dashed] (\R, 0) -- (\R, 1.5);
    \draw[dashed] (\R+\h, 0) -- (\R+\h, 1.5);
    \draw[<->] (\R, 1.2) -- (\R+\h, 1.2) node[midway, above] {$h$};
    
    \node at (\R+\h, 0) [below right] {$\theta=0$};

    \draw[->, thick] (O) -- (45:2) node[midway, above left] {$\mathbf{r}$};
    \draw[dashed] (O) -- (C);
    \draw (0.8,0) arc (0:45:0.8);
    \node at (1.1, 0.3) {$\theta$};

    \draw[->] (O) -- (\d+1,0) node[right] {$z$ (axis)};

    \shade[ball color=red] (C) circle (0.3);
    \node at (C) [right=0.3cm, above=0.3cm] {$M_c$};
    
\end{tikzpicture}
   \caption{Geometry of the tidal interaction. The primary star (left) is deformed by the companion (right). The tidal height $h$ is the difference in radius between the substellar point ($\theta=0$) and the pole ($\theta=\pi/2$).}
   \label{fig:tidal_geometry}
\end{figure}
\FloatBarrier

\noindent where $\theta$ is the angle between the position vector $\mathbf{r}$ and the direction of the companion.

We assume the atmosphere is in hydrostatic equilibrium. The deformation $h$ (the tidal height) is defined as the difference in the equipotential surface height between the point facing the companion ($\theta = 0$, the substellar point) and the plane perpendicular to the star-companion axis ($\theta = \pi/2$).

The difference in tidal potential between these two points at the stellar surface ($r=R_\star$) is:

\begin{equation}
   \Delta \Phi = \Phi_{\mathrm{tide}}(\pi/2) - \Phi_{\mathrm{tide}}(0) = \frac{G M_c R_\star^2}{2 d^3} - \left( -\frac{G M_c R_\star^2}{d^3} \right) = \frac{3 G M_c R_\star^2}{2 d^3} \enspace,
\end{equation}

To maintain an equipotential surface, this potential difference must be compensated by the work done against the surface gravity of the primary star, $g = G M_\star / R_\star^2$. Thus, $g h \approx \Delta \Phi$, which yields:

\begin{equation}
   h = \frac{\Delta \Phi}{g} = \left( \frac{3 G M_c R_\star^2}{2 d^3} \right) \left( \frac{R_\star^2}{G M_\star} \right) \enspace,
\end{equation}

\noindent Simplifying this expression leads to the scale height of the tide:

\begin{equation}
   h = \frac{3 M_c R_\star^4}{2 M_\star d^3} \enspace.
\end{equation}

\section{Uncertainties of time-averaged quantities}
\label{app:errors}

Time-averaged quantities from the simulations (face luminosities, their ratios, the centre-of-mass position) are computed from time series that are strongly correlated, because of convection and pulsation. The standard deviation $\sigma$ of the series overestimates the uncertainty of the mean, while $\sigma/\sqrt{N}$ underestimates it, where $N$ is the number of snapshots. We describe here how we estimate it.

\subsection{Autocorrelation time and effective sample size}

For a series $x_i$ ($i=1,\dots,N$) with mean $\bar{x}$, the normalised autocorrelation function is
\begin{equation}
\rho_k = \frac{\sum_{i=1}^{N-k} (x_i-\bar{x})(x_{i+k}-\bar{x})}{\sum_{i=1}^{N} (x_i-\bar{x})^2}.
\label{eq:acf}
\end{equation}
The integrated autocorrelation time, in units of snapshots, is
\begin{equation}
\tau = 1 + 2\sum_{k=1}^{M} \rho_k ,
\label{eq:tau}
\end{equation}
where the upper limit $M$ is chosen automatically as the smallest lag with $M > c\,\tau(M)$ and $c=5$ \citep{Sokal1997}. The number of statistically independent samples is
\begin{equation}
N_{\rm eff} = \frac{N}{\tau},
\label{eq:neff}
\end{equation}
and the uncertainty of a simple mean is
\begin{equation}
\sigma_{\bar{x}} = \frac{\sigma}{\sqrt{N_{\rm eff}}} = \sigma\sqrt{\frac{\tau}{N}} .
\label{eq:sem}
\end{equation}
For uncorrelated data, $\tau=1$ and Eq.~\eqref{eq:sem} reduces to $\sigma/\sqrt{N}$. For the luminosity we compute $\tau$ for the total luminosity and for the sum of the two faces of each axis, and use the largest value. The estimate is considered unreliable if $N < 50\,\tau$.

\subsection{Block bootstrap}

Eq.~\eqref{eq:sem} applies to a mean, but not to ratios (e.g.\ $L_x/\langle L_y, L_z\rangle$) or asymmetries. For these we use a block bootstrap \citep{Kunsch1989}, which keeps the correlation inside each block:
\begin{enumerate}
\item The series (all six faces together) is cut into $n_b = \lfloor N/\ell \rfloor$ consecutive blocks of length
\begin{equation}
\ell = \min\left(\lceil 5\,\tau \rceil,\; \lfloor N/10 \rfloor\right),
\label{eq:block}
\end{equation}
so that blocks are longer than the correlation time and there are at least $10$ of them.
\item $n_b$ blocks are drawn with replacement and concatenated into a resampled series of the same length.
\item The statistic $S$ (mean, ratio, asymmetry) is recomputed for the resampled series.
\item Steps 2--3 are repeated $B=2000$ times. The uncertainty is
\begin{equation}
\sigma_S^{\rm boot} = \sqrt{\frac{1}{B-1}\sum_{b=1}^{B}\left(S_b - \bar{S}_{\rm boot}\right)^2},
\label{eq:boot}
\end{equation}
where $\bar{S}_{\rm boot}$ is the mean of the $S_b$.
\end{enumerate}
The face luminosities are resampled together, so the correlations between faces are preserved in the ratios. For a mean, $\sigma_S^{\rm boot}$ approaches Eq.~\eqref{eq:sem}.

\subsection{Segment-to-segment check and adopted uncertainty}

As an independent estimate, each series is divided into $n_s=4$ consecutive segments, the statistic $S_j$ is computed in each, and
\begin{equation}
\sigma_S^{\rm seg} = \frac{1}{\sqrt{n_s}}\sqrt{\frac{1}{n_s-1}\sum_{j=1}^{n_s}\left(S_j-\bar{S}\right)^2}.
\label{eq:seg}
\end{equation}
This estimate is sensitive to variations slower than $\ell$, which the bootstrap can underestimate when the run is short compared to $\tau$. We adopt the more conservative value,
\begin{equation}
\sigma_S = \max\left(\sigma_S^{\rm boot},\,\sigma_S^{\rm seg}\right).
\label{eq:adopted}
\end{equation}
The significance of a difference between a model and the single-star reference is $z = (S-S_{\rm ref})/\sqrt{\sigma_S^2+\sigma_{S,{\rm ref}}^2}$.

\subsection{Limitations}

Both methods assume a stationary series. With only a few correlation times in a run, $\tau$ itself is uncertain, which is why we keep the segment estimate, and the uncertainties should be seen as approximate.





\FloatBarrier
\onecolumn
\clearpage
\section{Figures}

\begin{figure}[h!]
   \centering
   \includegraphics[width = \textwidth]{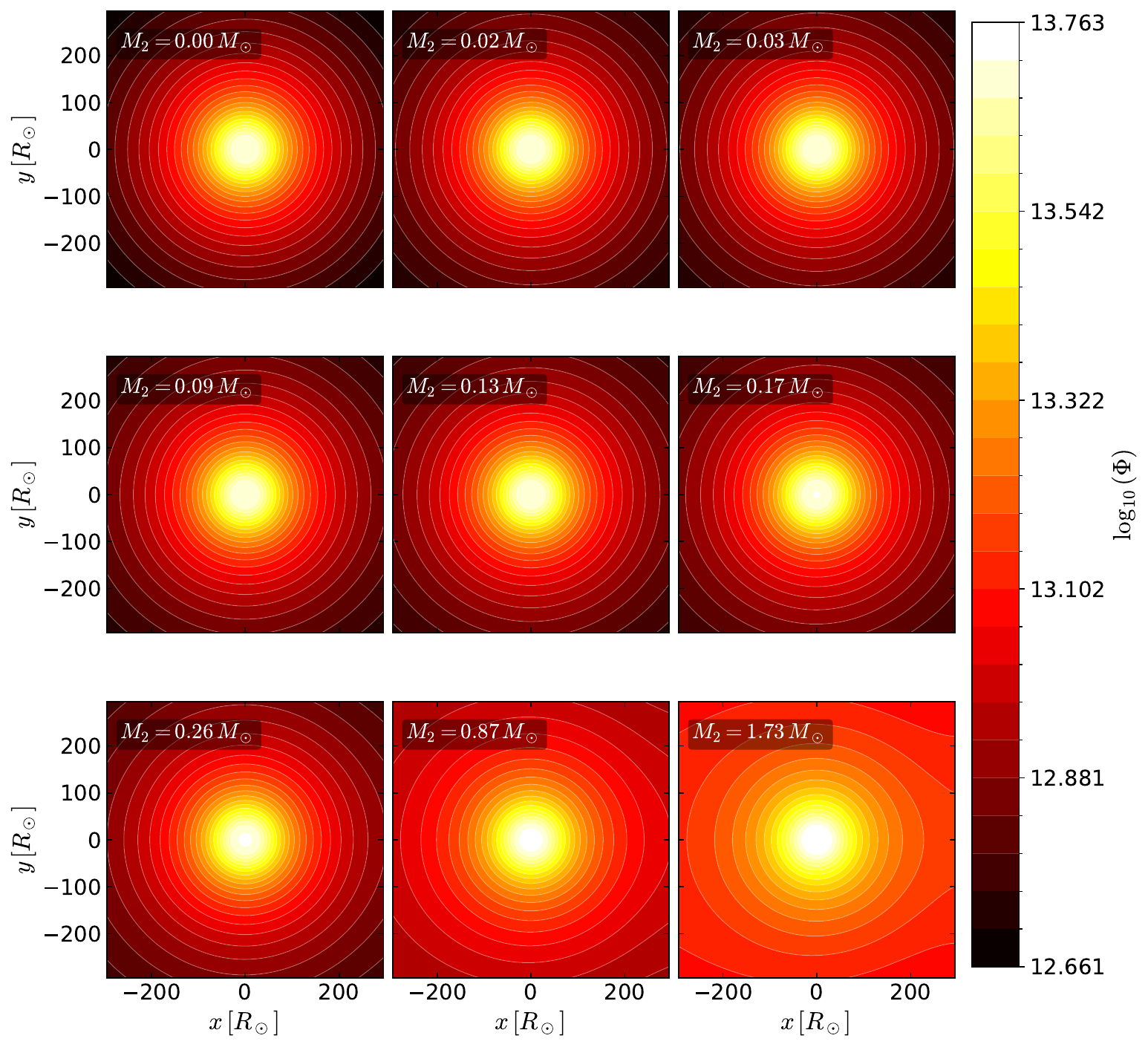}
   \caption{Gravitational potential $\Phi(\mathbf{r})$ for different companion masses $M_c$. The primary star is at the origin, and the companion is located at $d = 3.34$\,AU along the x-axis. The color scale represents the potential depth, with lighter regions indicating stronger gravitational influence.}
\end{figure}

\FloatBarrier

\begin{figure}
   \centering
   \subfloat{\includegraphics[width=\textwidth]
   {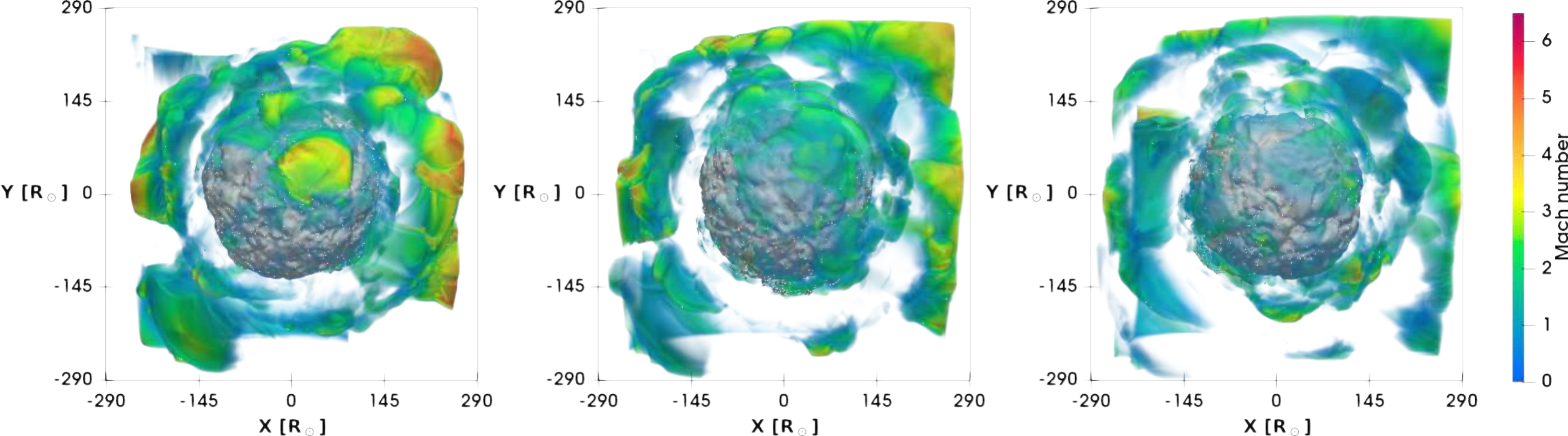}}\\
   \subfloat{\includegraphics[width=\textwidth]{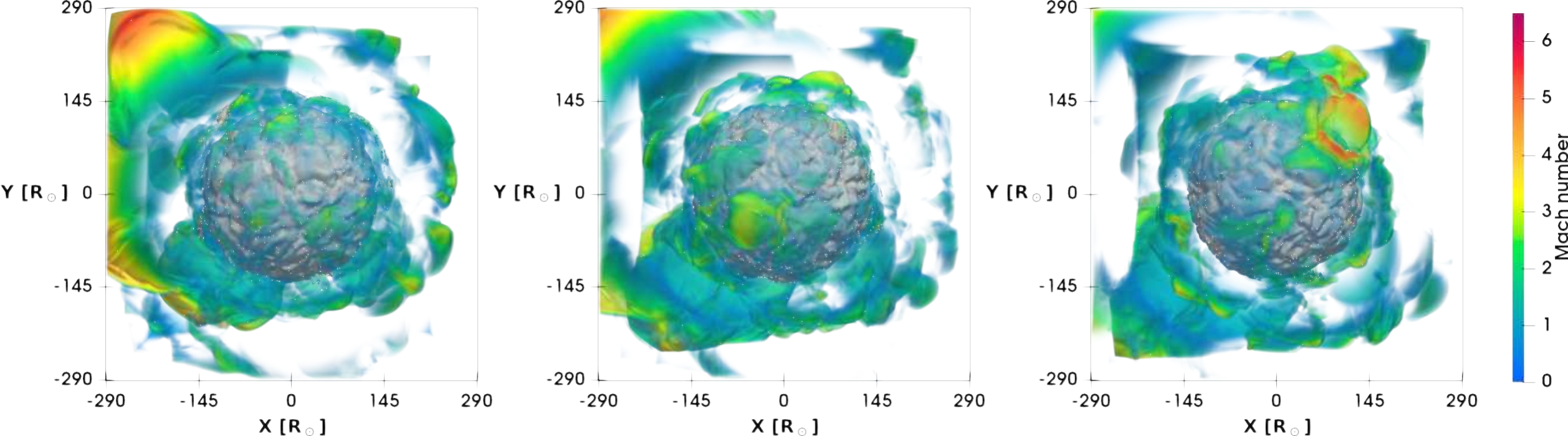}}\\
   \subfloat{\includegraphics[width=\textwidth]{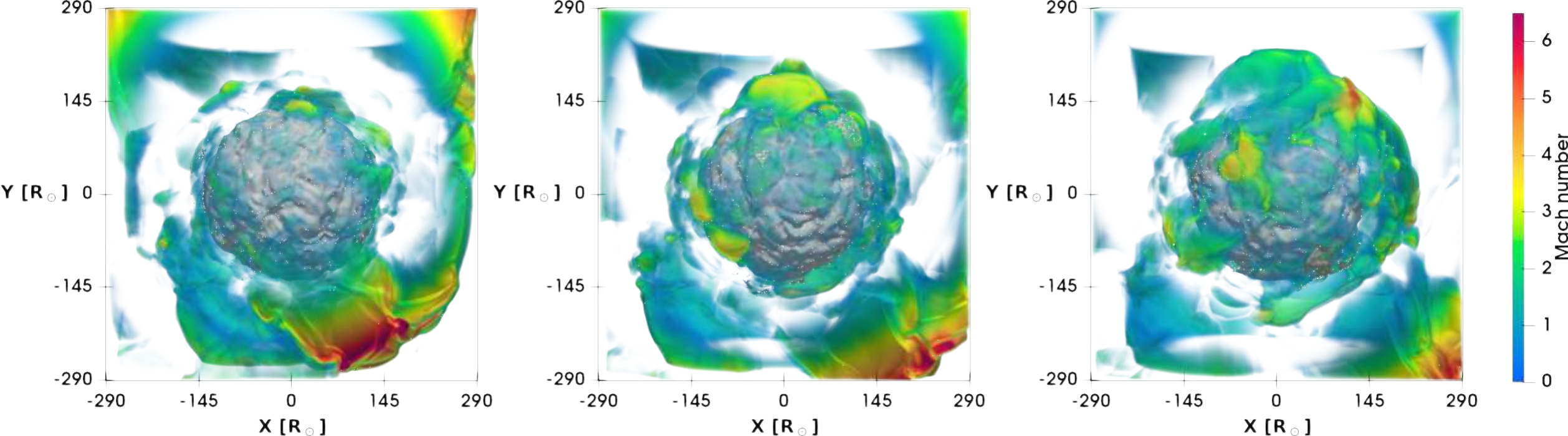}}\\
   \subfloat{\includegraphics[width=\textwidth]{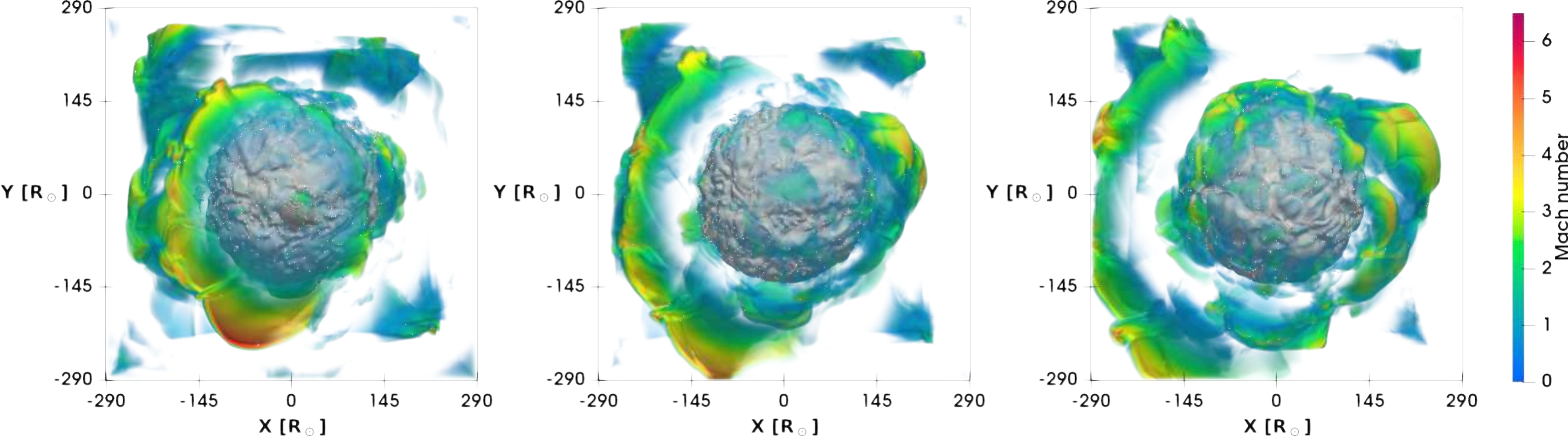}}\\
   \caption{3D snapshots of the Mach number for each model. The color scale represents the Mach number, with warmer colors indicating higher speeds. The grey sphere represents the isothermal surface at $T = 4000$\,K, which corresponds to the temperature where $\tau = 33$. Each snapshot is spaced by ~ 1 year.}
   \label{fig:snapshots}
\end{figure}

\begin{figure}
   \ContinuedFloat
   \centering
   \subfloat{\includegraphics[width=\textwidth]{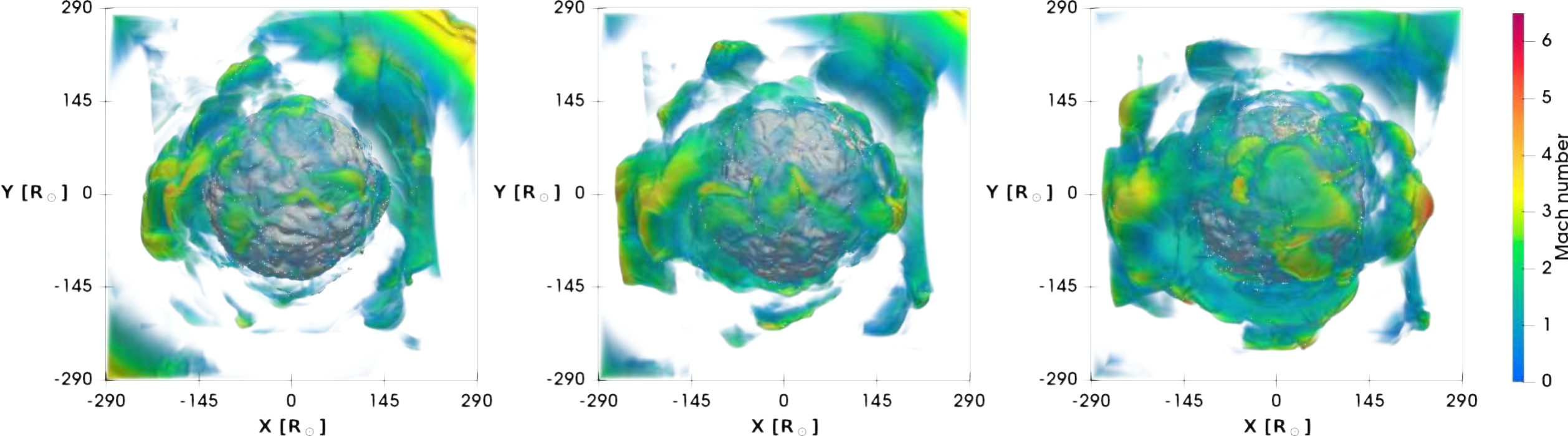}}\\
   \subfloat{\includegraphics[width=\textwidth]{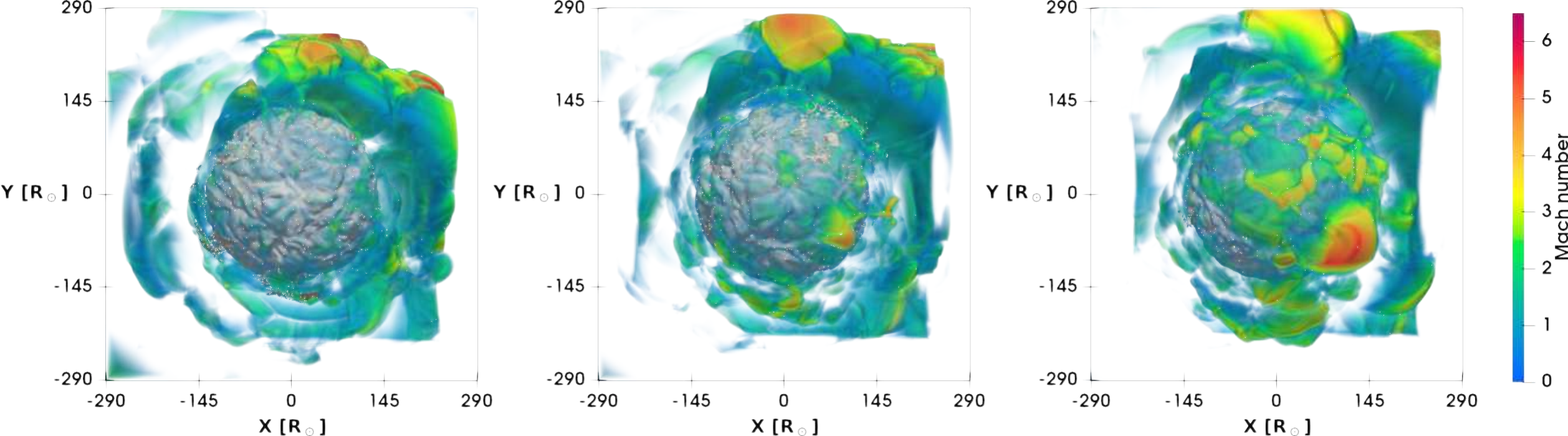}}\\
   \subfloat{\includegraphics[width=\textwidth]{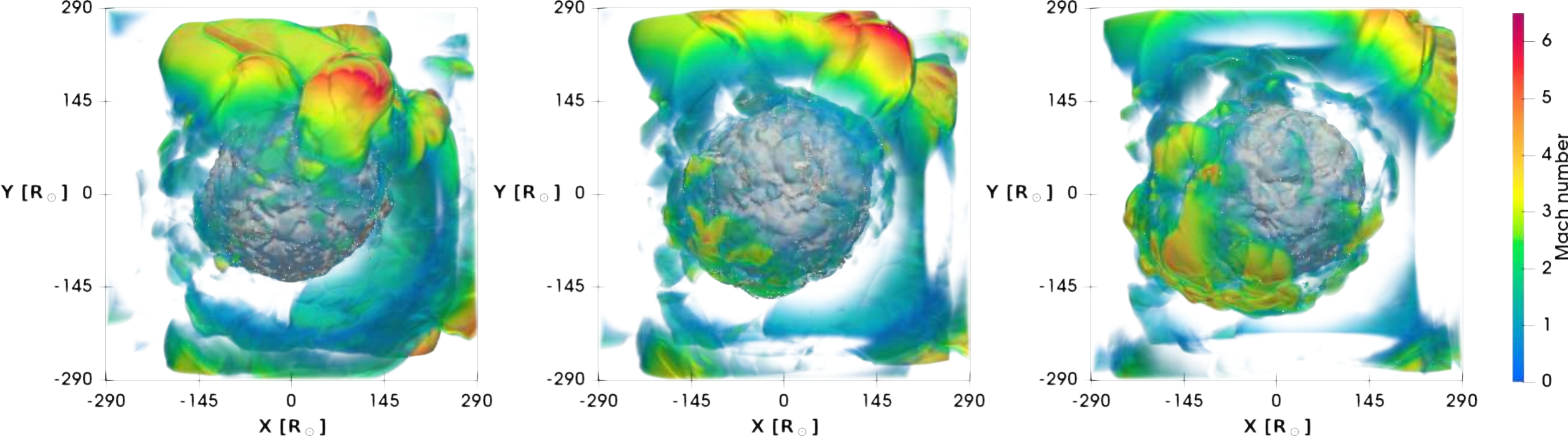}}\\
   \subfloat{\includegraphics[width=\textwidth]{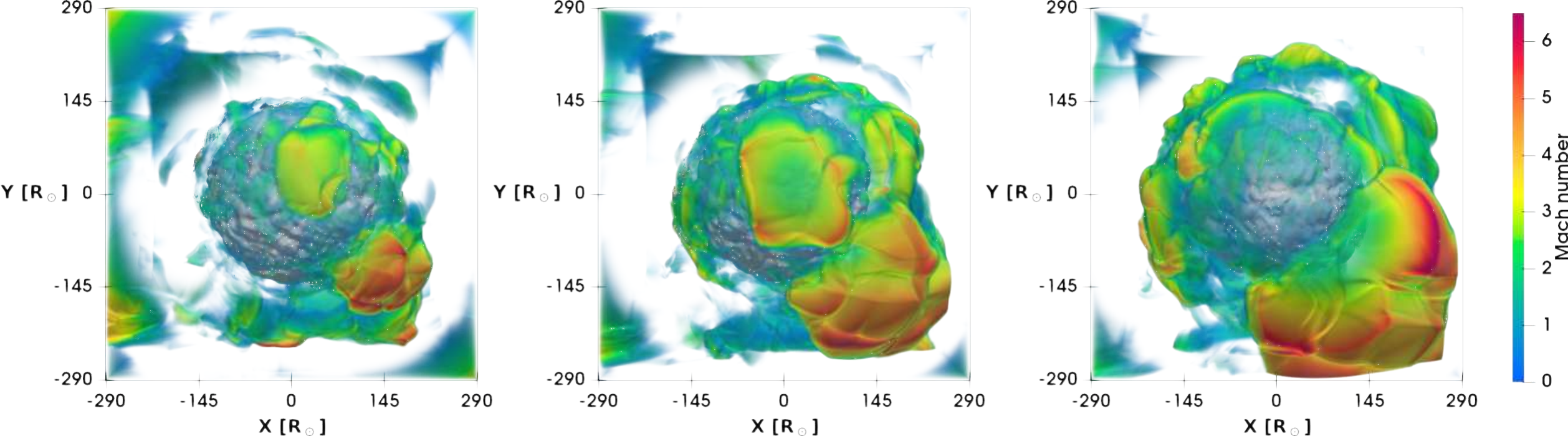}}\\
   \caption{\textit{(continued.)}}
\end{figure}

\begin{figure}
   \ContinuedFloat
   \centering
   \subfloat{\includegraphics[width=\textwidth]{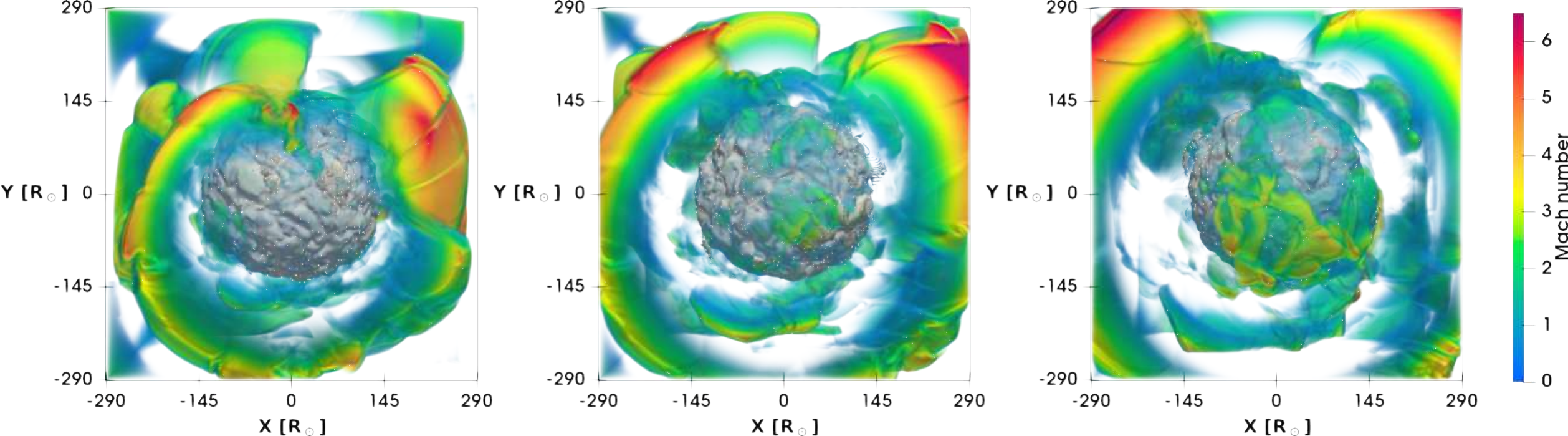}}\\
   \caption{\textit{(continued.)}}
\end{figure}

\end{appendix}
\end{document}